\documentclass[iop,revtex4]{emulateapj}

\usepackage{subfigure}

\usepackage{lscape}
\usepackage{natbib}

\usepackage{txfonts}
\usepackage{hyperref}       
\usepackage{epsfig}

\usepackage{multirow}

\providecommand{\abs}[1]{\mid#1\mid}

\renewcommand{\anchor}[2]{\href{#1}{#2}}

\newcommand{\htwo}  {H~{\sc ii}}
\newcommand{\htr}  {{\htwo\ region}}
\newcommand{\htrs}  {{\htwo\ regions}}

\newcommand{\es}  {{erg s$^{-1}$}}

\defcitealias{johnson06-12}{J12}
\defcitealias{hwang04-08}{HL08}

\newcommand{\HST}{{\em Hubble Space Telescope}}
\newcommand{\HSTt}{{\em HST}}

\newcommand{\Chandra}{{\em Chandra}}
\newcommand{\Chandras}{{\em Chandra's}}

\newcommand{\xmmn}{{\em XMM-Newton}}
\newcommand{\rosat}{{\em ROSAT}}

\newcommand{\ACIS}   {{ACIS}}
\newcommand{\CIAO}   {{\em CIAO}}

\newcommand{\CALDB} {{\em CALDB}}

\journalinfo{Accepted by: $\textit{The Astrophysical Journal}$}

\shorttitle{Faint X-Ray Binaries in M31}
\shortauthors{Vulic et. al.}

\begin{document}

\title{Faint X-Ray Binaries and Their Optical Counterparts in M31}

\author{N. Vulic\altaffilmark{1}\altaffilmark{$\bigstar$}, S. C. Gallagher\altaffilmark{1}, and P. Barmby\altaffilmark{1}}

\altaffiltext{1}{Department of Physics \& Astronomy, Western University, London, ON, N6A 3K7, Canada}

\altaffiltext{$\bigstar$}{nvulic@uwo.ca}

\begin{abstract}

X-ray binaries (XRBs) are probes of both star formation and stellar mass, but more importantly remain one of the only direct tracers of the compact object population. 
To investigate the XRB population in M31, we utilized all 121 publicly available observations of M31 totalling over 1 Ms from \emph{Chandra's} ACIS instrument. We studied 83 star clusters in the bulge using the year 1 star cluster catalogue from the Panchromatic Hubble Andromeda Treasury Survey. We found 15 unique star clusters that matched to 17 X-ray point sources within 1\arcsec\ (3.8 pc). This population is composed predominantly of globular cluster low-mass XRBs, with one previously unidentified star cluster X-ray source. 
Star clusters that were brighter and more compact preferentially hosted an X-ray source. Specifically, logistic regression showed that the F475W magnitude was the most important predictor followed by the effective radius, while color (F475W$-$F814W) was not statistically significant. We also completed a matching analysis of 1566 \htrs\ and found 10 unique matches to 9 X-ray point sources within 3\arcsec\ (11 pc). The \htrs\ hosting X-ray point sources were on average more compact than unmatched \htrs, but logistic regression concluded that neither the radius nor H$\alpha$ luminosity was a significant predictor. Four matches have no previous classification and thus are high-mass XRB candidates. A stacking analysis of both star clusters and \htrs\ resulted in non-detections, giving typical upper limits of $\approx10^{32}$ \es, which probes the quiescent XRB regime.

\end{abstract}

\section{Introduction} \label{sec:intro}

\setcounter{footnote}{1}

Compact objects are the end-states of the evolution of massive stars and as such are signposts of the star formation history of a galaxy. X-ray binaries (XRBs) consist of a compact object, either a neutron star or black hole, which accretes matter from a companion star. Active XRBs have luminosities of $\sim$10$^{35-41}$ \es\ (those $>$$10^{39}$ \es\ are classified as ultra-luminous X-ray sources) while those in quiescence are $\lesssim$10$^{34}$ \es. XRBs are classified into two main categories based on the mass of the companion star: low-mass (LMXB) and high-mass (HMXB) \citep{fabbiano09-06}. LMXBs accrete matter via Roche lobe overflow while HMXBs transfer mass predominantly by Bondi-Hoyle (wind) accretion \citep{bondi-44, iben09-95}. LMXB formation is more efficient in globular clusters (GCs) than in the field of a galaxy due to the higher stellar densities in GCs \citep{katz02-75, clark08-75, fabian08-75, pooley07-03}. In the Milky Way, LMXB formation is approximately two orders of magnitude more efficient in GCs \citep{katz02-75, clark08-75}. Conversely, HMXBs are associated with star-forming regions (OB associations, \htwo\ regions, and infrared-bright dusty regions) as opposed to star clusters \citep{ranalli02-03, grimm03-03, swartz10-04, persic02-07, shtykovskiy05-07, lehmer11-10, walton09-11, swartz11-11, mineo01-12}. Due to their evolutionary timescales, LMXBs trace the stellar mass of a galaxy \citep{gilfanov03-04, kim08-04, zhang09-11} while HMXBs probe the star formation rate \citep{grimm03-03, mineo01-12} within the past $\sim$100 Myr \citep{shtykovskiy05-07}. 

The Milky Way's XRBs have been studied extensively in the Galactic Center, star clusters, and the field \citep[e.g.][]{grimm09-02, revnivtsev11-08, muno03-09, bodaghee01-12, lutovinov05-13, nebot-gomez-moran05-13}. Faint X-ray sources (quiescent XRBs, cataclysmic variables, millisecond pulsars, etc.) in the Milky Way have been studied to 10$^{30}$ \es\ and lower \citep{heinke11-03, sazonov04-06, heinke11-06} as a result of \Chandras\ subarcsecond resolution and optical counterpart identifications with the \HST\ (\HSTt). Due to dust obscuration, we can only probe X-ray sources in the plane of the disk out to $\approx$8 kpc from the Solar System. This hinders our ability to obtain a complete sample of the Galaxy's XRB population. Fortunately, M31 provides a convenient nearby analogue that allows us to perform an analysis of the faint X-ray population in a large galaxy.  

M31's XRB population has been studied numerous times over the past decades with various X-ray telescopes ($\emph{X-ray Multi-Mirror Mission}$ (\xmmn), $\emph{R\"ontgensatellit}$ (\rosat), $\emph{Einstein}$). 
However, only the spatial resolution of \Chandra\ allows us to study low-luminosity X-ray sources by separating them from the diffuse emission in the disk and crowding in the bulge.
The LMXBs in M31 have been extensively studied by various groups \citep{trudolyubov12-04, voss06-07, voss10-07, peacock10-10, zhang09-11, barnard09-122, barnard09-12, barnard06-13} that report on X-ray luminosity functions, spatial distributions, variability, and spectral analysis. \citet{peacock10-10} used \xmmn\ to study 80\% of the 416 confirmed GCs in M31 \citep{peacock02-10}. They found 41 GCs associated with X-ray sources along with an additional 4 GCs identified with \Chandra\ and \rosat, for a total of 11\% of GCs with X-ray sources. M31's GCs with LMXBs were found to be brighter, redder (metal-rich), and more compact than GCs without LMXBs, in agreement with other extragalactic XRB studies \citep{sivakoff05-07, paolillo08-11, mineo01-14}. 
\citet{voss10-07} predicted a large number of faint transient sources in the bulge of M31, similar to the numerous faint X-ray sources (accreting millisecond pulsars) found in the Galactic Center \citep{muno03-09}. 

A recent study by \citet{stiele10-11} with \xmmn\ covered all of M31 for the first time and reported on the X-ray source population compared to all previous studies. The survey catalogued 1897 point sources (914 new) and reached a limit of $\sim$10$^{35}$ \es\ in the $0.2-4.5$ keV energy band. Sources were classified/identified by X-ray hardness ratios, spatial extent and distribution, cross-correlations in other wavelengths, and variability (using \Chandra\ and \rosat\ observations). They found 10 field LMXBs and 26 field LMXB candidates by analyzing long-term X-ray source variability. In addition, another 36 LMXBs were associated with GCs and 17 LMXB candidates were associated with GC candidates. No HMXBs were identified - in fact none have been confirmed in M31 to date - although 2 new candidates were proposed in addition to the 18 candidates from \citet{shaw-greening03-09}. The comprehensive X-ray population analysis by \citet{stiele10-11} has classified hundreds of sources and shed light on their long-term variability and spatial/flux distributions. However, they pointed out that $\sim$65\% of their point sources are classified as ``hard'', meaning that they have no optical identification and their X-ray colors are ambiguous: they could be XRBs, supernova remnants, or background active galaxies.

\citet{barnard01-14} have attempted to address the issue of unclassified X-ray point sources by identifying them using X-ray data alone. They used structure functions, which estimate the mean intensity deviation of data over a time interval. By determining structure functions for each source they constrained its variability and were able to differentiate XRBs from active galactic nuclei. They found 220 X-ray sources above $\gtrsim$10$^{35}$ \es\ with significantly more variability than expected in the structure function for active galactic nuclei. Based on their analysis they classified these sources as XRBs, with an additional 30 XRB sources for a total of 250 probable XRBs (200 new) out of their sample of 528 X-ray sources in the central 20\arcmin\ of M31. Low-luminosity XRBs are more variable than luminous XRBs and thus are well-suited to this classification technique given sufficient signal-to-noise data. 

One of the most powerful methods to confidently classify an X-ray source is identifying an optical counterpart. However, in crowded and extragalactic fields this generally requires the exquisite spatial resolution of \HSTt. The ability to observe a large fraction of identified X-ray sources is limited since \HSTt\ has both high demand and a small field of view. 
The Panchromatic $\emph{Hubble}$ Andromeda Treasury (PHAT) survey \citep{dalcanton06-12} is a multicycle \HSTt\ program to map one-third of M31's disk and provide an unprecedented catalogue of star clusters and stars. While it does not cover all of M31, it provides the best possible resource for determining optical counterparts to X-ray sources on such a large scale. M31's bulge has been monitored extensively with \Chandra, whose 0.5\arcsec\ point spread function (PSF) is superior to the 6\arcsec\ provided by \xmmn.

To probe low-luminosity X-ray sources in M31 below the level achieved by \citet{stiele10-11}, we will use the stacking method developed by \citet{brandt07-01, brandt09-01} and \citet{hornschemeier06-01}. By stacking star cluster positions obtained from the PHAT survey and \htwo\ region positions (see Section \ref{sec:opt}) in a \Chandra\ X-ray image of M31, we can determine, on average, if these objects host faint point sources. Young star clusters and \htwo\ regions will probe the HMXB population while GCs will probe the LMXBs. Our study is not focused on the field population of XRBs. Stacking analyses enable us to study the effect that metallicity has on faint X-ray sources by separating metal-rich and metal-poor classes of GCs. Stacking is a unique way to investigate whether HMXBs in M31 exist but are below our current detection limits. By pushing extragalactic XRB luminosities into the quiescent regime ($<$10$^{34}$ \es), we can reveal undetected populations that will help constrain binary evolution models \citep[e.g.][]{hurley02-02, kiel07-06, belczynski01-08, lipunov10-09, bhadkamkar02-12, bhadkamkar01-132, siess02-13, bhadkamkar04-14}.

We adopt a distance to M31 of 776 $\pm$ 18 kpc as in \citet[hereafter J12]{johnson06-12}, which corresponds to a linear scale of 3.8 pc arcsecond$^{-1}$. In Section \ref{sec:obs}, we describe the optical and X-ray observations and reduction procedure for the X-ray data. In Section \ref{sec:match}, we report results from matching star clusters and \htrs\ to X-ray point sources. In Section \ref{sec:stacking}, we present results from an X-ray stacking analysis of star clusters and \htrs. In Sections \ref{sec:discuss} and \ref{sec:summary} we discuss our results and summarize our main conclusions.

\section{Observations} \label{sec:obs}

\subsection{X-ray Data}	\label{sec:xrdata}

\begin{figure*}[!ht]
\plotone{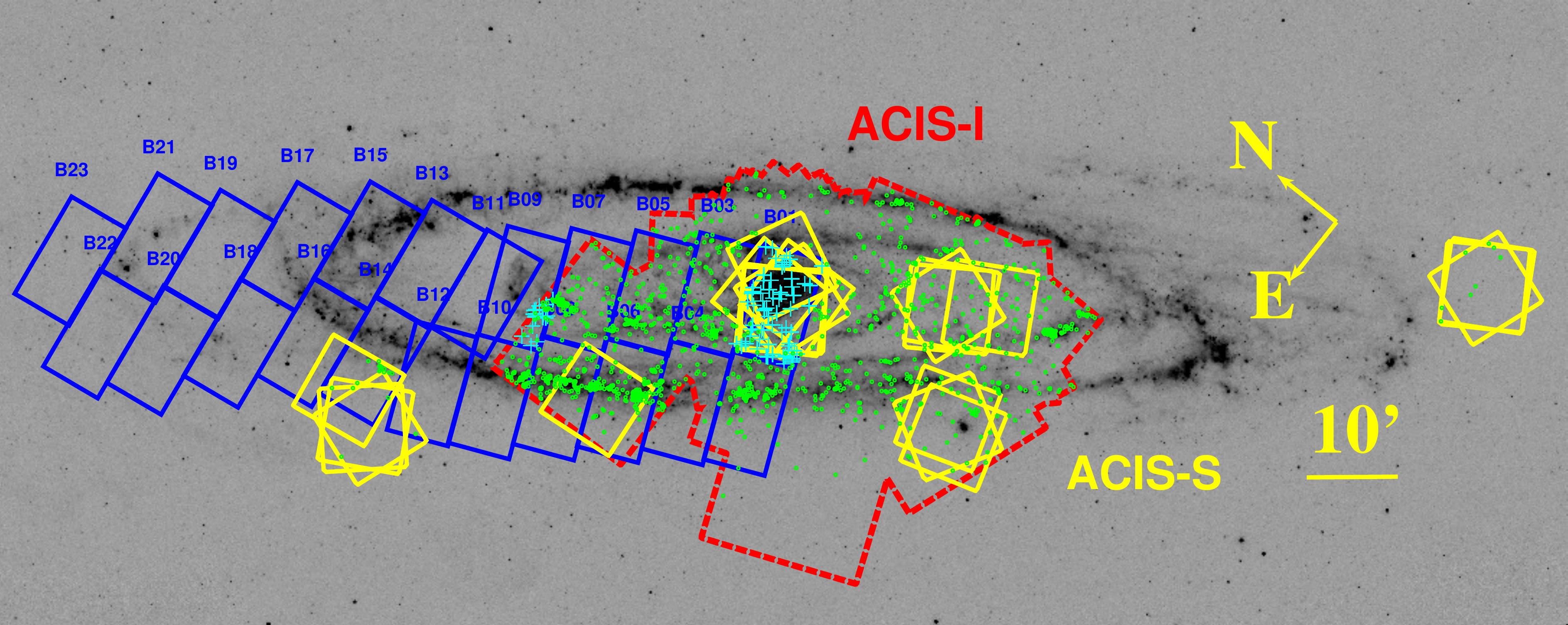}

\caption{The field of view of the ACIS-I (dashed outline, red in the online version) merged observations and ACIS-S3 chips (yellow in the online version) for each observation used in our analysis overlaid on a Spitzer 24 $\micron$ image of M31 \citep{gordon02-06}. The PHAT footprint is included with each rectangular brick (blue in the online version) labelled to show the overlap with \Chandra\ data. The year 1 PHAT star cluster catalogue only includes bricks 1, 9, 15, 21, and parts of 17 and 23. Crosses (cyan in the online version) and circles (green in the online version) represent star clusters and \htrs, respectively, within the field of view of \Chandra\ data (see Section \ref{sec:opt} for more details). Most \Chandra\ observations were part of an ongoing monitoring program of the supermassive black hole, while off-center ACIS-S pointings studied supersoft X-ray sources.}
\label{fig:acis-fov}
\end{figure*}

\Chandra~has observed M31 numerous times, but the small field of view (compared to \xmmn) of $\emph{Chandra's}$ Advanced CCD Imaging Spectrometer (\ACIS) instrument ($8\arcmin \times 8\arcmin$ for ACIS-S and $16\arcmin \times 16\arcmin$ for ACIS-I) combined with the proximity of M31 require many pointings to cover an appreciable area of the galaxy. We use all 121 publicly available ACIS-S and ACIS-I observations in M31 for our matching and stacking analyses of star clusters and \htwo\ regions, having a total exposure time over 1 Ms. Due to the degradation of the PSF at large off-axis angles, we only use data from the S3 chip from the ACIS-S observations. The observations are listed in Tables \ref{tab:acis-s} and \ref{tab:acis-i} and their fields of view shown in Figure \ref{fig:acis-fov}.

\begin{deluxetable*}{cccccccc}
\tablewidth{0pc}
\tablecolumns{8}
\tablecaption{ACIS-S Observations\label{tab:acis-s}}
\tablehead{
\colhead{ObsID\tablenotemark{a}} & \colhead{Date} & \colhead{R.A.} & \colhead{Decl.} & \colhead{Livetime} & \colhead{Datamode} & \colhead{CCDs} & \colhead{Roll Angle} \\
\colhead{} & \colhead{} & \colhead{(J2000)} &  \colhead{(J2000)} &  \colhead{(s)} &  \colhead{} &  \colhead{} &  \colhead{(degrees)} }
\startdata
\dataset[ADS/Sa.CXO#obs/00309]{309} & 2000-06-01 & 10.688 & 41.27877 & 5129 & FAINT & 235678 & 87.5\\
\dataset[ADS/Sa.CXO#obs/00310]{310} & 2000-07-02 & 10.68311 & 41.27876 & 5284 & FAINT & 235678 & 108.7\\
\dataset[ADS/Sa.CXO#obs/00313]{313} & 2000-09-21 & 10.65816 & 40.87837 & 6002 & FAINT & 235678 & 162.9\\
\enddata
\tablecomments{Table 1 is published in its entirety in the electronic edition of the Astrophysical Journal. A portion is shown here for guidance regarding its form and content. Positions represent the aimpoint.}
\tablenotetext{a}{An asterisk indicates an interleaved observation.}
\end{deluxetable*}

\begin{deluxetable*}{cccccccc}
\tablewidth{0pc}
\tablecolumns{8}
\tablecaption{ACIS-I Observations\label{tab:acis-i}}
\tablehead{
\colhead{ObsID\tablenotemark{a}} & \colhead{Date} & \colhead{R.A.} & \colhead{Decl.} & \colhead{Livetime} & \colhead{Datamode} & \colhead{CCDs} & \colhead{Roll Angle} \\
\colhead{} & \colhead{} & \colhead{(J2000)} &  \colhead{(J2000)} &  \colhead{(s)} &  \colhead{} &  \colhead{} &  \colhead{(degrees)} }
\startdata
\dataset[ADS/Sa.CXO#obs/00303]{303} & 1999-10-13 & 10.67806 & 41.26989 & 11890 & FAINT & 012367 & 193.0\\
\dataset[ADS/Sa.CXO#obs/00305]{305} & 1999-12-11 & 10.68273 & 41.26395 & 4185 & FAINT & 01236 & 274.0\\
\dataset[ADS/Sa.CXO#obs/00306]{306} & 1999-12-27 & 10.68409 & 41.2637 & 4189 & FAINT & 01236 & 285.4\\
\enddata
\tablecomments{Table 2 is published in its entirety in the electronic edition of the Astrophysical Journal. A portion is shown here for guidance regarding its form and content. Positions represent the aimpoint.}
\tablenotetext{a}{An asterisk indicates an interleaved observation.}
\end{deluxetable*}

\subsubsection{Data Reduction} \label{sec:xraydata}

Each observation was reprocessed with the \Chandra~Interactive Analysis of Observations (\CIAO) tools package version 4.5 \citep{fruscione07-06} and the \Chandra~Calibration database (\CALDB) version 4.5.5 \citep{graessle07-06}. 
We started with the level-1 events file and produced a bad pixel file that identified observation-specific bad pixels, hot pixels, bright bias pixels, and afterglow events using $\texttt{acis\_run\textunderscore hotpix}$. Since we are studying faint sources, we ran $\texttt{acis\_detect\_afterglow}$ to eliminate cosmic ray afterglows with only a few events. 
The level-1 events file was updated and filtered using the standard (ASCA) grades ($0, 2-4, 6$), status bits ($0$), good time intervals, charge transfer inefficiency (CTI), time-dependent gain, and pulse height with $\texttt{acis\_process\_events}$. We used the energy-dependent subpixel event repositioning algorithm to improve astrometry and, where necessary, set the VFAINT option in $\texttt{acis\_process\_events}$. 

We then created merged event files for all 26 ACIS-S observations and all 95 ACIS-I observations with $\texttt{reproject\_obs}$ to obtain fully-reduced images in the soft, hard, and full-bands corresponding to energy ranges of $0.3-2$ keV, $2-8$ keV, and $0.3-8$ keV respectively. 
We produced exposure-corrected images (in units of photons cm$^{-2}$ s$^{-1}$ pixel$^{-1}$) with the $\texttt{flux\_obs}$ tool using a binsize of 1 to maintain the native resolution. Weighted spectrum files (corresponding to the soft, hard, and full energy bands) were used to calculate instrument maps. The merged ACIS-S and ACIS-I images have dimensions\footnote{Since \Chandra\ ACIS-S observations are sparse (Figure \ref{fig:acis-fov}) large parts of the image are devoid of data.} of 1.7$\degr$ $\times$ 1.7$\degr$ and 0.8$\degr$  $\times$ 0.85$\degr$ and correspond to exposures of $\sim$548 ks and $\sim$ 510 ks respectively\footnote{These exposure times are not consistent across the fields of view since merging results in overlap of varying exposures from each observation.}. 
The plate scale of the \Chandra\ images is 0.5 arcsecond pixel$^{-1}$, corresponding to 1.9 pc pixel$^{-1}$. \Chandras\ absolute astrometry is $\sim$0.6\arcsec\ at the 90\% confidence level \citep{broos05-10} and has spatial resolution that ranges from 1\arcsec\ on-axis to 4\arcsec\ at 4\arcmin\ off-axis for 1.5 keV at 90\% encircled energy fraction for ACIS\footnote{\url{http://cxc.harvard.edu/proposer/POG/}}.

\subsubsection{X-Ray Point Source Detection} \label{sec:xrdata-detect}

Given that we have produced deep images of M31's bulge, there may be new point sources that have not been previously identified. We used $\texttt{wavdetect}$ on the full, hard, and soft-band images with scales\footnote{Using the $\sqrt{2}$ series from 1 to 8 we detected an additional 60 X-ray sources. However, this does not change the results of our matching analysis (see Section \ref{sec:res-matching}).} of 1, 2, 4, 8, and 16 as recommended by the Chandra X-ray Center to have a wide range of source sizes\footnote{\url{http://cxc.harvard.edu/ciao/ahelp/wavdetect.html\#plist.scales}}. This ensures that the variation of the PSF at different off-axis angles is accounted for. Exposure maps were also input to reduce false positives. The sigthresh parameter, which determines how many false detections are allowed, should not be larger than the inverse of the number of pixels in the image being analyzed. Therefore it was set to $3\times10^{-8}$ for the ACIS-I merged image and $7\times10^{-9}$ for the ACIS-S merged image. These numbers are small since the ACIS-I merged image (0.68 deg$^{2}$) covers an area larger than the bulge and the ACIS-S merged image (2.89 deg$^{2}$) covers the majority of the galaxy even though only the nucleus and various pointings have data (see Figure \ref{fig:acis-fov} and Section \ref{sec:xraydata}). In any case, setting the sigthresh parameter to the default value of $10^{-6}$ did not change the number of X-ray sources matched to star clusters or \htrs. We obtained source lists for the full, hard, and soft-band merged ACIS-I and ACIS-S images. From the six source lists we collated our results into a master source list with the $\texttt{match\_xy}$ tool from the Tools for ACIS Review and Analysis (TARA) package\footnote{\url{http://www2.astro.psu.edu/xray/docs/TARA/}}. We found a total of 1033 independent X-ray point sources across all 3 bands (a source was not necessarily detected in more than one band) for the ACIS-I merged observations and ACIS-S merged observations. We report the number of sources detected in each energy band for both ACIS-I and ACIS-S in Table \ref{tab:xray-detect}. While the $\texttt{match\_xy}$ tool filtered X-ray sources based on their positional uncertainties from $\texttt{wavdetect}$ (e.g.\ retains only one source position if detected in all 3 energy bands), $\texttt{match\_xy}$ does not consolidate pairs of extremely close $\texttt{wavdetect}$ sources detected only in one band. We performed an internal match of our 1033 sources to remove sources within $0.5\arcsec$ of another source. This found a further 112 matches that we removed to obtain 921 effective X-ray sources.

\begin{deluxetable}{ c c  c  c  c c }
\tabletypesize{\scriptsize}
\tablecaption{X-ray Sources Detected in Merged ACIS Observations \label{tab:xray-detect}}
\tablecolumns{6}
\tablewidth{0pt}
\tablehead{
\colhead{Detector}	&
\colhead{Full}	&		
\colhead{Hard}	&
\colhead{Soft}	&
\colhead{Total}	&
\colhead{Total Effective Sources}
}
\startdata
ACIS-I	&	435	&	344	&	388	& 1167	&	\multirow{2}{*}{921}
\\	ACIS-S	&	353	&	217	&	334	& 904	&
\\
\enddata
\tablecomments{Total Effective Sources (ACIS-I \& ACIS-S) refers to the number of actual X-ray sources detected after filtering those that appear in multiple energy bands and on both detectors. An internal match to within $0.5\arcsec$ filtered out sources with separations closer than the pixel scale.} 
\end{deluxetable}

We input our list of sources to the ACIS Extract package \citep{broos05-10} along with exposure maps (calculated using weighted spectrum files) so that each observation could be individually analyzed. ACIS Extract then merged all observations (ACIS-I and ACIS-S) and produced the relevant source properties and statistics, such as X-ray luminosity. ACIS Extract aligns each observation to a published astrometric catalogue\footnote{Naval Observatory Merged Astrometric Dataset \citep{zacharias05-04} or 2MASS Point Source Catalog \citep{skrutskie02-06}} and thus gives precision of $\sim$0.1\arcsec\ for most point sources. We required ACIS Extract's PROB\_NO\_SOURCE statistic for each X-ray source to be $>99.9\%$, except where by-eye inspection of the images showed source crowding that biased this statistic to lower values (i.e.\ 5 sources included that were not $>99.9\%$). 
The majority of our sources ($\sim$75\%) are located in the bulge, where the \Chandra\ observations are deeper compared with the rest of the galaxy. We only processed (through ACIS Extract) the 26 X-ray sources that matched to star clusters and \htrs\ (see Section \ref{sec:res-matching} for details). The luminosity range of these 26 X-ray sources was $8.5\times10^{33}$ \es\ to $7.8\times10^{37}$ \es. There is no meaningful, single value for the completeness limit since the exposure time varies significantly across the usable field of view. The deepest exposure of any 1 pixel reaches $\sim500$ ks in both the merged ACIS-I and merged ACIS-S images; this occurs in the nucleus and corresponds to a sensitivity limit of $\approx10^{33}$ \es. The complete X-ray point source catalogue will follow in a subsequent paper (Vulic et al., in prep.).

\subsection{Optical Data} \label{sec:opt}

We use the star cluster catalogue compiled by \citetalias{johnson06-12} from the PHAT survey. The survey operates using the Advanced Camera for Surveys (filters F475W and F814W) as well as the Wide Field Camera 3 (filters F275W, F336W, F110W, and F160W) on the \HSTt. The observational technique used by the survey is detailed in \citet{dalcanton06-12} and comprises 23 areas known as ``bricks'', with each one consisting of 18 different fields of view. The survey produces photometry with a signal-to-noise ratio of 4 at $m_{F275W}$ = 25.1, $m_{F336W}$ = 24.9, $m_{F475W}$ = 27.9, $m_{F814W}$ = 27.1, $m_{F110W}$ = 25.5, and $m_{F160W}$ = 24.6 for single pointings in the uncrowded outer disk. The PHAT year 1 star cluster catalogue was compiled from bricks 1, 9, 15, 21, and parts of 17 and 23. Due to the limited \Chandra\ data we only utilize brick 1 and a small part of brick 9 (proposal 12058, PI: Julianne Dalcanton). Even though Figure \ref{fig:acis-fov} shows numerous PHAT bricks coincident with \Chandra\ data, the only bricks with published star cluster catalogues that overlap are 1 and 9.  There are 83 star clusters\footnote{A PHAT year 1 `possible cluster' catalogue exists that we have not used.} in the field of view of the \Chandra\ observations that we use for our matching and stacking analyses (indicated by crosses in Figure \ref{fig:acis-fov}). \citetalias{johnson06-12} have corrected the absolute astrometry to agree with the 2MASS reference system within $\sim$60 mas. The PSF (full width at half maximum)  for F475W is $\approx0.10\arcsec-0.13\arcsec$ \citep{ubeda12-12}.

In addition, we use the new catalogue of \htrs\ in M31 compiled by \citet{azimlu10-11}. M31's entire disk is covered out to $\sim$24 kpc from the center based on observations with the Mayall 4 m telescope as part of the Nearby Galaxies Survey \citep{massey05-06}. A total of 3691 \htwo\ regions down to L$_{\rm{H}\alpha} = 10^{34}$ \es\ are identified in the 2.2 deg$^{2}$ mosaic, with 1566 in the field of view of the \Chandra\ data (indicated by circles in Figure \ref{fig:acis-fov}). The average PSF (full width at half maximum) of the survey was 1\arcsec, which is generally better than \Chandras\ range of $1\arcsec-4\arcsec$ but not as good as the $\approx0.10\arcsec-0.13\arcsec$ for F475W from the PHAT survey. The astrometric calibration of the \htr\ catalogue is good to $0.1\arcsec$, better than \Chandras\ 0.6\arcsec\ but not as precise as the 60 mas for PHAT.

\section{Data Analysis}	\label{sec:match}

\subsection{Matching Results} \label{sec:res-matching}

\begin{figure*}[!ht]
\epsscale{0.50}
\begin{center}
\begin{tabular}{cc}
\plotone{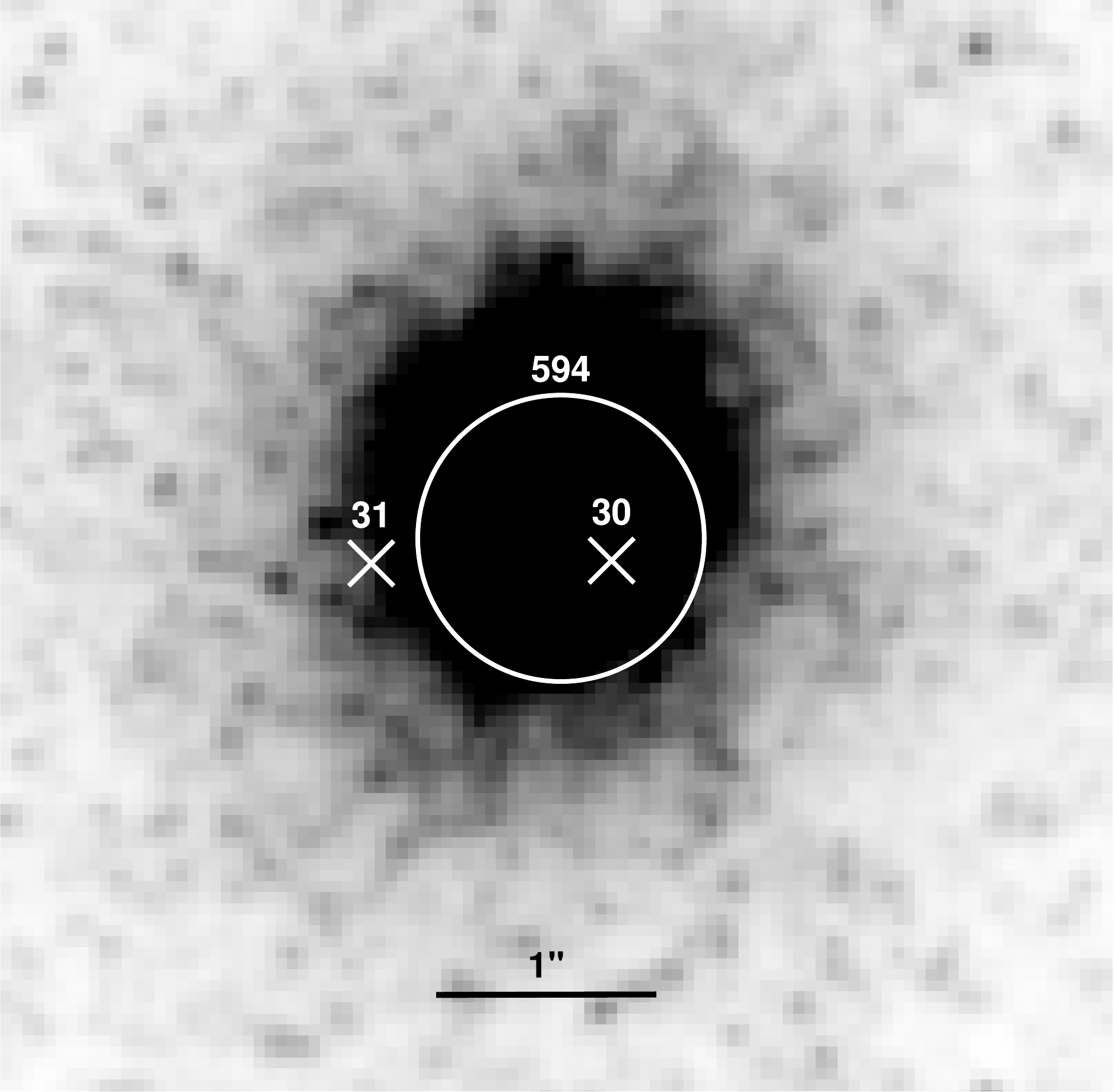}
\plotone{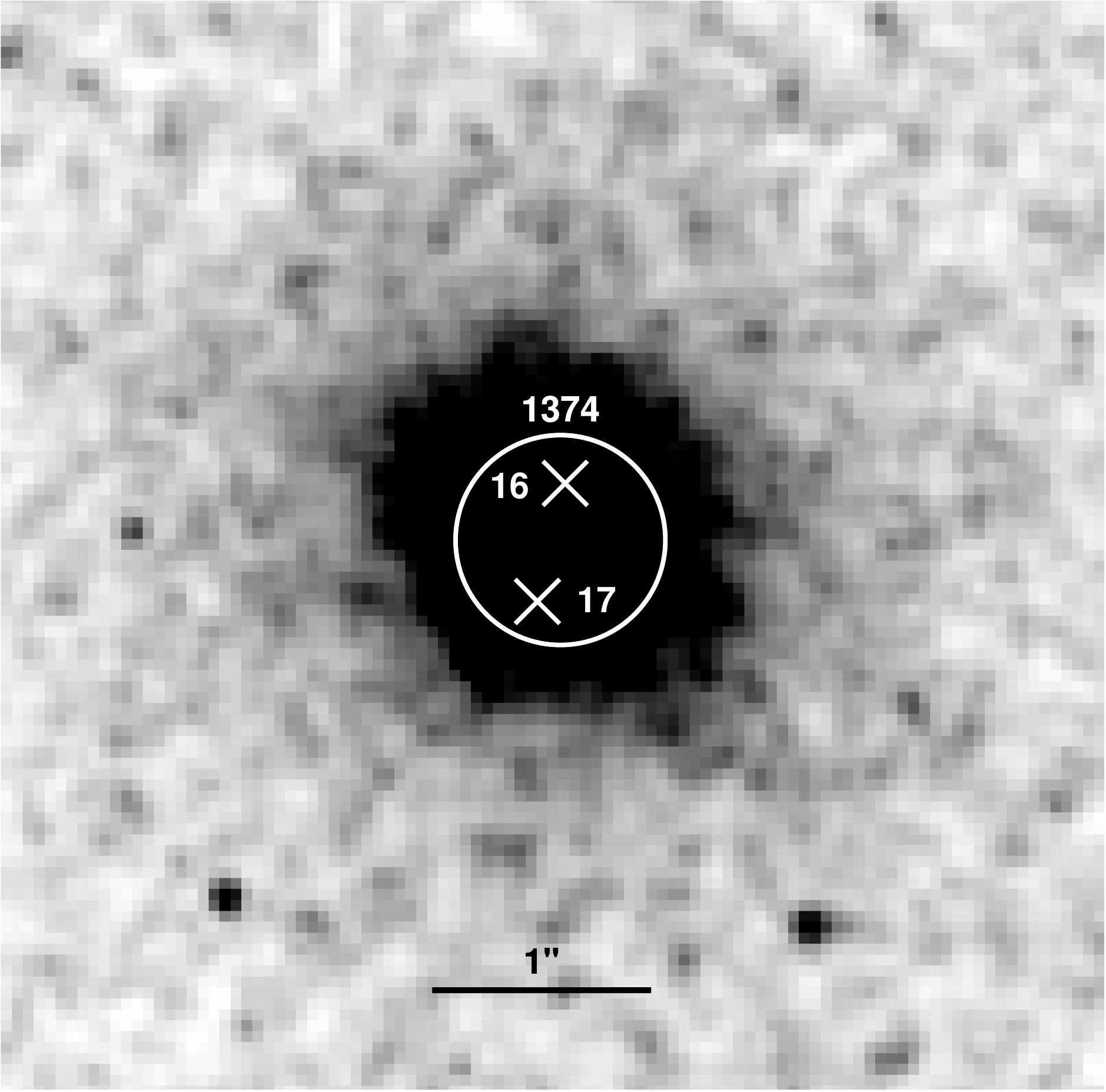}
\end{tabular}
\caption{Images from the F475W band of the PHAT survey showing star clusters with double X-ray point source matches, where each panel is 5\arcsec\ on a side. White circles represent star clusters (with the radius equal to the effective radius of the star cluster) while an ``$\times$'' represents the position of an X-ray source. The numbers correspond to the PHAT and AE identification numbers respectively from Table \ref{tab:scxrmatches}.} 
\label{fig:doubles}
\end{center}
\end{figure*}

Prior to stacking star clusters and \htwo\ regions we must exclude those with identified X-ray point sources. Stacking probes faint X-ray point sources by co-adding star clusters (or \htrs) that are undetected in X-rays. In this way one hopes to detect a signal in the source aperture (centered on the position of the star cluster or \htr) above the background (see Section \ref{sec:stacking} for more details). Therefore, including a star cluster (or \htr) with a detected X-ray point source in the complete stack of all undetected star clusters (or \htrs) would defeat the purpose of detecting a faint X-ray source in the stack. Figure \ref{fig:acis-fov} shows where the star clusters and \htrs\ we analyzed are located in M31. The 83 star clusters in the field of view of \Chandra\ data are located mainly in the bulge (PHAT brick 1), while a handful are in the disk (PHAT brick 9); the 1566 \htrs\ are spread throughout the disk. Here we summarize our matching results, beginning with star clusters.

We used TOPCAT \citep{taylor12-05} to complete a matching analysis with the 83 star clusters and 921 X-ray point sources. A total of 15 star clusters were matched to 17 unique X-ray point sources within 1$\arcsec$. The matching results are shown in Table \ref{tab:scxrmatches}. Two star clusters (PHAT IDs 594 and 1374) each had two X-ray point source matches. Figure \ref{fig:doubles} shows these star clusters in the F475W band image from PHAT. The clusters are plausible double XRB hosts since the X-ray sources are separated by $>$0.5\arcsec. Even though we used a 1\arcsec\ matching radius, only 2 star clusters (PHAT IDs 1403 and 1415) had a separation $>$0.5\arcsec\ from their matched X-ray point source. No additional clusters were matched when increasing the matching radius to $>$10\arcsec. 
Only one star cluster (PHAT ID 1374) has not previously been associated with an X-ray source. 
In addition, one star cluster (PHAT ID 1375) that was previously identified as a GC candidate X-ray source by \citep{voss06-07} was not associated with an X-ray source in our analysis. Upon further inspection of our ACIS-I image it seemed that $\texttt{wavdetect}$ did not identify what was likely an X-ray source.

To address the chance coincidence probability of our matches we created 25 source lists from our original X-ray catalogue that offset each source by $\pm$ 5$-$10$\arcsec$ \citep{antoniou06-09, zezas10-02}. A false-match rate of $\sim$2\% was recovered ($\leq$1 false match). We also used the cumulative number counts of \citet{brandt12-01} to estimate the expected background sources in the 83 star cluster regions of radius 1\arcsec\ each. In equation \ref{eq:bkgrnd}, $S$ is the flux in \es\ cm$^{-2}$ and $N$ is the number of background sources per deg$^{2}$ above that flux, where we used the faintest X-ray point source from our matches ($\approx1.4\times10^{-16}$ \es\ cm$^{-2}$). This gave $N(>1.4\times10^{-16}$ \es\ cm$^{-2}$) = $(9.4\times10^{-4})^{+5.4\times10^{-4}}_{-1.3\times10^{-3}}$ expected background sources, indicating that our matches are likely authentic. We also completed cone searches in the NASA/IPAC Extragalactic Database (NED) within $30\arcsec$ around each of the 17 X-ray point sources that were matched to star clusters and found no background galaxies. 
\begin{eqnarray}
N(>S) & = & 3970\left(\frac{S}{1\times10^{-16}}\right)^{-0.67\pm 0.14} \rm{deg^{-2}}	\label{eq:bkgrnd}
\end{eqnarray}

\begin{deluxetable*}{  c c  c  c  c  c  c  c  c  c  c  c  c  c  c  c  }
\tabletypesize{\scriptsize}
\tablecaption{Star Cluster \& X-ray Point Source Matches \label{tab:scxrmatches}}
\tablecolumns{16}
\tablewidth{0pt}
\tablehead{
\multicolumn{3}{c|}{Star Cluster Positions\tablenotemark{a}} & \multicolumn{8}{c|}{ Results from ACIS Extract} & \multicolumn{2}{c}{Group Data\tablenotemark{b}} & \colhead{} & \multicolumn{2}{c}{Previous Classification} \\
\cline{1-3} \cline{4-11} \cline{12-13} \cline{15-16} \\
\colhead{PHAT ID} &
\colhead{RA} &
\colhead{Dec} &
\colhead{AE ID} &
\colhead{RA} &
\colhead{Dec} &
\colhead{PErr\tablenotemark{c}} &
\colhead{Prob\tablenotemark{d}} &
\colhead{OAA\tablenotemark{e}} &
\colhead{PSF \%} &
\colhead{$\log(L_{X})\tablenotemark{f}$} &
\colhead{ID} &
\colhead{Size} &
\colhead{Sep} &
\colhead{Class\tablenotemark{g}} &
\colhead{Ref.\tablenotemark{h}} \\
\colhead{} &
\multicolumn{2}{c}{(J2000)} &
\colhead{} &
\multicolumn{2}{c}{(J2000)} &
\colhead{($\arcsec$)} &
\colhead{} &
\colhead{($\arcmin$)} &
\colhead{} &
\colhead{(\es)} &
\colhead{} &
\colhead{} &
\colhead{($\arcsec$)} &
\colhead{} &
\colhead{}
}
\startdata
  594 & 10.856666 & 41.260284 & 30 & 10.856581 & 41.260256 & 0.00 & 1.00\tablenotemark{i}  & 4.3 & 0.4 & 33.32 & 1 & 2 & 0.25 &  GC  &  HO 
 \\ 594 & 10.856666 & 41.260284 & 31 & 10.856984 & 41.260252 & 0.00 & 1.00\tablenotemark{i}  & 4.3 & 0.4 & 33.37 & 1 & 2 & 0.87 &  GC  &  HO
  \\ 1374 & 10.711676 & 41.285381 & 16 & 10.711668 & 41.285453 & 0.15 & 0.11\tablenotemark{j}  & 1.6 & 0.39 & 34.24 & 2 & 2 & 0.26 &  \nodata  &  \nodata 
 \\ 1374 & 10.711676 & 41.285381 & 17 & 10.711715 & 41.285303 & 0.25 & 0.70\tablenotemark{j}  & 1.6 & 0.39 & 27.88\tablenotemark{k} & 2 & 2 & 0.30 &  \nodata  &  \nodata  
  \\ 612 & 10.749429 & 41.268238 & 21 & 10.749429 & 41.268268 & 0.01 & 0.00 & 3.5 & 0.9 & 37.61 &  \nodata  &  \nodata  & 0.11 &  GC/BH?  &  ZH/BA 
  \\ 616 & 10.766111 & 41.301309 & 24 & 10.766068 & 41.301343 & 0.01 & 0.00 & 2.0 & 0.39 & 37.42 & \nodata & \nodata & 0.17 &  GC/BH?  &  ZH/BA 

 \\ 642 & 10.794234 & 41.247575 & 27 & 10.794231 & 41.2476 & 0.02 & 0.00 & 3.1 & 0.39 & 37.82 & \nodata & \nodata & 0.09 &  GC/BH?  &  ZH/BA 

 \\ 663 & 10.762226 & 41.25627 & 23 & 10.7622 & 41.256281 & 0.01 & 0.00 & 3.2 & 0.4 & 37.37 & \nodata & \nodata & 0.08 &  GC/BH?  &  VO/BA 
  \\ 682 & 10.814255 & 41.190284 & 28 & 10.814253 & 41.190228 & 0.13 & 0.00 & 7.0 & 0.9 & 35.98 &  \nodata  &  \nodata  & 0.20 &  GC  &  VO 
 \\ 1377 & 10.731713 & 41.309748 & 19 & 10.731691 & 41.309742 & 0.02 & 0.00 & 3.1 & 0.39 & 36.39 & \nodata & \nodata & 0.06 &  fgStar/GC  &  VO/KO 
 \\ 1381 & 10.692025 & 41.293409 & 13 & 10.692013 & 41.293426 & 0.03 & 0.00 & 1.5 & 0.39 & 35.91 & \nodata & \nodata & 0.07 &  GC  &  KO 
 \\ 1386 & 10.63027 & 41.327441 & 6 & 10.630254 & 41.327479 & 0.01 & 0.00 & 4.2 & 0.4 & 37.24 & \nodata & \nodata & 0.14 &  GC  &  VO 
 \\ 1395 & 10.672678 & 41.256623 & 10 & 10.672652 & 41.256609 & 0.01 & 0.00 & 1.0 & 0.39 & 36.51 & \nodata & \nodata & 0.09 &  GC  &  VO 
  \\ 1401 & 10.638559 & 41.295108 & 7 & 10.638544 & 41.295104 & 0.02 & 0.00 & 2.6 & 0.9 & 36.31 &  \nodata  &  \nodata  & 0.04 &  GC  &  HO 
 \\ 1403 & 10.623894 & 41.2993 & 4 & 10.624093 & 41.299152 & 0.39 & 0.00 & 3.5 & 0.9 & 34.79 &  \nodata  &  \nodata  & 0.76 &  GC  &  ZH 
 \\ 1415 & 10.68207 & 41.211915 & 11 & 10.681932 & 41.211796 & 0.28 & 0.07 & 3.7 & 0.9 & 33.82 &  \nodata  &  \nodata  & 0.57 &  GC  &  HO
 \\ 1444 & 10.58984 & 41.238797 & 3 & 10.589851 & 41.238801 & 0.20 & 0.00 & 4.2 & 0.4 & 35.85 & \nodata & \nodata & 0.03 &  IRsrc  &  KA 
 \\
\enddata
\tablenotetext{a}{Taken from the catalogue of \citetalias{johnson06-12}.}
\tablenotetext{b}{Group ID identifies star clusters that were matched to multiple X-ray point sources (Group Size).}
\tablenotetext{c}{Source positional uncertainty.}
\tablenotetext{d}{Probability source does not exist.}
\tablenotetext{e}{Average off-axis angle.}
\tablenotetext{f}{In the 0.3 $-$ 8 keV band.}
\tablenotetext{g}{BH = Black Hole, GC = globular cluster, fgStar = foreground star, IRsrc = infrared source, ? indicates a candidate.}
\tablenotetext{h}{BA \citep{barnard06-13}, HO \citep{hofmann07-13}, KA \citep{kaaret10-02}, KO \citep{kong10-02}, VO \citep{voss06-07}, ZH \citep{zhang09-11}.}
\tablenotetext{i}{X-ray sources located at the edge of ACIS-S image and thus only contribute to the background, resulting in a large probability.}
\tablenotetext{j}{Suffer from crowding and thus have unreasonably high values.}
\tablenotetext{k}{Source has a large $A_{V}$ from best-fit XSPEC \citep{arnaud96} model and thus the luminosity is unreliable.}
\end{deluxetable*}

Our results show that $\sim$18\% (15/83) of star clusters were matched to an X-ray point source within 1\arcsec. In a survey of eight Fornax and Virgo elliptical galaxies (limiting luminosities typically few $\times10^{37}$ \es) \citet{kim02-13} found that 408 of 5904 ($\sim$7\%) GCs are associated with an X-ray source within 0.5\arcsec. Similarly, a survey of 11 Virgo Cluster early-type galaxies (limiting luminosities typically few $\times10^{37}$ \es) by \citet{sivakoff05-07} found that 270 of 6488 ($\sim$4\%) GCs were associated with an X-ray source within 1\arcsec. Our analysis was restricted to the bulge/central region of M31 and thus the concentration of X-ray point sources is larger than throughout the rest of the galaxy (as \citealt{muno03-09} found for the Milky Way). Since LMXBs trace stellar mass \citep{gilfanov03-04}, and stellar mass density is larger in the bulge, we would therefore expect to have a higher percentage of matches (LMXBs) compared to the value for the entire galaxy. Additionally, our luminosity limit is a factor of $10^{3}$ fainter than previous surveys and therefore increases our percentage of matches. To more directly compare different galaxies, we compare the fraction (and number) of GCs with an LMXB versus those without above a given luminosity limit from our survey with those of \citet{sivakoff05-07} and \citet{kim02-13} in Table \ref{tab:xrb-compare}. We compared our results to individual galaxies from these two surveys since most galaxies they studied had varying limiting $L_{X}$ (i.e.\ their complete sample is not completeness-corrected). The fractions are comparable (e.g.\ 6\% for our work and 6.5\% for \citealt{sivakoff05-07}) at low-$L_{X}$ limits ($L_{X}>5\times10^{37}$ \es). For $L_{X}>2.6\times10^{37}$ \es\ a fraction of 3.6\% found in our study is much lower than the 8.3\% of \citet{kim02-13}. In general, we and \citet{sivakoff05-07} find a difference in the fraction that is lower by a factor of $2-4$ than \citet{kim02-13} for $L_{X}>2.6\times10^{37}$ \es, which could be attributed to small-number statistics. When analyzing the same galaxies as \citet{sivakoff05-07}, \citet{kim02-13} found more GC-LMXBs as a result of deeper X-ray observations over an extended period of time, which also allowed for the identification of previously quiescent sources that went into outburst.

Of the 1566 \htwo\ regions in the field of view, we found 10 that matched to 9 unique X-ray point sources within 3\arcsec\ ($\sim$11.2 pc, their average radius). The results are shown in Table \ref{tab:htwoxrmatches}. The false-match rate for \htrs\ was $\sim$35\%, which is higher due to the increased matching radius and their larger physical size compared to star clusters. For each of the 9 X-ray point sources that were matched to \htrs, we completed NED cone searches using a radius of 30\arcsec. Only one source (AE ID 38, 2.3\arcsec\ from an \htr; see Table \ref{tab:htwoxrmatches}) had any nearby entries in NED. These were a nova at a distance of 0.4\arcsec, an \htr\ at a distance of 3.2\arcsec, and a background galaxy at a distance of 7\arcsec. We summarize our matching results and state completeness limits in Table \ref{tab:match-results}.

\begin{deluxetable*}{ c c c  c c c c}[!h]
\tabletypesize{\scriptsize}
\tablecaption{GC-LMXB Fraction Comparisons \label{tab:xrb-compare}}
\tablecolumns{7}
\tablewidth{0pt}
\tablehead{
\colhead{}		&
\colhead{}		&
\multicolumn{5}{c}{Percentage of GCs with an LMXB}  \\
\colhead{Study}		&
\colhead{Galaxies}		&
\colhead{$\gtrsim0.05$}	&
\colhead{$\gtrsim0.09$} &
\colhead{$\gtrsim0.14$}	&
\colhead{$\gtrsim0.26$}	&
\colhead{$\gtrsim0.3$} \\
\colhead{}		&
\colhead{}		&
\multicolumn{5}{c}{($10^{38}$ erg s$^{-1}$)}
}
\startdata
	This work & M31 & 6\% (5) & 6\% (5) & 6\% (5) & 3.6\% (3) & 2.4\% (2)
\\ \hline
\\	\multirow{2}{*}{\citet{kim02-13}\tablenotemark{a}} & NGC 4365, NGC 4649 & \nodata & \nodata &  \nodata & 8.3\% (209) & \nodata
\\	 & NGC 4472 &  \nodata & \nodata & \nodata & \nodata & 9.2\% (70)
\\ \hline
\\	\multirow{4}{*}{\citet{sivakoff05-07}\tablenotemark{b}} & NGC 4374, NGC 4697 & 6.5\% (52) & \nodata & \nodata & \nodata & \nodata
\\	 & NGC 4365, NGC 4406 & \nodata & 3.8\% (49) &  \nodata & \nodata & \nodata
\\	 & NGC 4486, NGC 4552 & \nodata & \nodata & 3.8\% (80) & \nodata & \nodata
\\	 & NGC 4526 & \nodata & \nodata & \nodata & \nodata & 2.9\% (7)
\\
\enddata
\tablecomments{Values shown in brackets represent the number of GC-LMXBs. Since the comparison studies do not publish values above a given $L_{X}$ for the entire sample we can only compare individual galaxies.}
\tablenotetext{a}{Sample of Virgo and Fornax cluster early-type galaxies; uses $L_{X,min}$ ($0.3-10$ keV) as limiting luminosity.}
\tablenotetext{b}{Sample of Virgo cluster early-type galaxies; uses $L_{X,min}$ ($0.3-8$ keV) at 50\% detection probability as limiting luminosity.}
\end{deluxetable*}

\begin{deluxetable*}{ c c  c  c  c  c  c  c  c  c  c  c  c  c  c  c  c  c }
\tabletypesize{\scriptsize}
\tablecaption{\htwo\ Region \& X-ray Point Source Matches \label{tab:htwoxrmatches}}
\tablecolumns{18}
\tablewidth{0pt}
\tablehead{
\multicolumn{5}{c|}{\htwo\ Region Data\tablenotemark{a}} & \multicolumn{8}{c|}{Results from ACIS Extract} & \multicolumn{2}{c}{Group Data\tablenotemark{b}} & \colhead{} & \multicolumn{2}{c}{Previous Classification} \\
\cline{1-5} \cline{6-13} \cline{14-15} \cline{17-18} \\
\colhead{H{\sc ii} ID} &
\colhead{RA} &
\colhead{Dec} &
\colhead{Radius} &
\colhead{$\log(L_{\rm{H}\alpha})$} &
\colhead{AE ID} &
\colhead{RA} &
\colhead{Dec} &
\colhead{PErr\tablenotemark{c}} &
\colhead{Prob\tablenotemark{d}} &
\colhead{OAA\tablenotemark{e}} &
\colhead{PSF \%} &
\colhead{$\log(L_{X})\tablenotemark{f}$} &
\colhead{ID} &
\colhead{Size} &
\colhead{Sep} &
\colhead{Class\tablenotemark{g}} &
\colhead{Ref.\tablenotemark{h}} \\
\colhead{} &
\multicolumn{2}{c}{(J2000)} &
\colhead{(pc)} &
\colhead{(\es)} &
\colhead{} &
 \multicolumn{2}{c}{(J2000)} &
\colhead{($\arcsec$)} &
\colhead{} &
\colhead{($\arcmin$)} &
\colhead{} &
\colhead{(\es)} &
\colhead{} &
\colhead{} &
\colhead{($\arcsec$)} &
\colhead{} &
\colhead{}
}
\startdata
2165 & 10.8667 & 41.3092 & 10.99 & 36.87 & 33 & 10.866947 & 41.309129 & 0.14 & 0.00 & 7.8 & 0.5 & 35.94 & 1 & 2 & 0.7 &  SNR  &  WI/ST 
 \\ 2164 & 10.8667 & 41.3091 & 10.43 & 36.94 & 33 & 10.866947 & 41.309129 & 0.14 & 0.00 & 7.8 & 0.5 & 35.94 & 1 & 2 & 0.7 &  SNR  &  WI/ST 
 \\ 1686 & 10.6429 & 40.9527 & 13.33 & 35.31 & 8 & 10.643727 & 40.95233 & 0.46 & 0.00 & 4.2 & 0.9 & 35.87 &  \nodata  &  \nodata  & 2.6 &  GC  &  PE/ZH/ST 
 \\ 1798 & 10.7046 & 41.4017 & 7.83 & 36.72 & 14 & 10.704479 & 41.402351 & 0.28 & 0.00 & 7.9 & 0.9 & 35.53 &  \nodata  &  \nodata  & 2.4 &  SNR  &  ST 
 \\ 1805 & 10.7083 & 40.8914 & 8.62 & 34.44 & 15 & 10.707519 & 40.891337 & 0.59 & 0.00 & 3.7 & 0.9 & 34.86 &  \nodata  &  \nodata  & 2.1 &  \nodata  &  \nodata 
 \\ 1831 & 10.7233 & 41.4307 & 9.48 & 36.91 & 18 & 10.722953 & 41.430795 & 0.31 & 0.00 & 9.3 & 0.9 & 35.99 &  \nodata  &  \nodata  & 1.0 &  SNR  &  ST/VO 
 \\ 2080 & 10.8287 & 41.338 & 7.83 & 35.64 & 29 & 10.828635 & 41.338083 & 0.25 & 0.00 & 7.2 & 0.9 & 35.61 &  \nodata  &  \nodata  & 0.3 &  SSS?  &  HO/ST 
 \\ 2241 & 10.885 & 41.3499 & 7.83 & 35.11 & 34 & 10.884631 & 41.349752 & 1.09 & 0.23 & 8.6 & 0.9 & 35.02 &  \nodata  &  \nodata  & 1.1 &  \nodata  &  \nodata 
 \\ 2471 & 10.9721 & 41.2017 & 10.43 & 36.03 & 36 & 10.972348 & 41.200971 & 0.62 & 0.00 & 6.9 & 0.9 & 35.44 &  \nodata  &  \nodata  & 2.7 &  \nodata  &  \nodata 
 \\ 3005 & 11.1367 & 41.4225 & 7.83 & 37.52 & 38 & 11.13587 & 41.422606 & 0.76 & 0.00 & 3.9 & 0.9 & 35.11 &  \nodata  &  \nodata  & 2.3 &  \nodata  &  \nodata 
 \\ 
\enddata
\tablenotetext{a}{Taken from the catalogue of \citet{azimlu10-11}.}
\tablenotetext{b}{Group ID identifies an \htr\ matched to multiple X-ray point sources (Group Size).}
\tablenotetext{c}{Source positional uncertainty.}
\tablenotetext{d}{Probability source does not exist.}
\tablenotetext{e}{Average off-axis angle.}
\tablenotetext{f}{In the 0.3 $-$ 8 keV band.}
\tablenotetext{g}{SNR = supernova remnant, GC = globular cluster, SSS = supersoft X-ray source, ? indicates a candidate.}
\tablenotetext{h}{WI \citep{williams07-04}, ST \citep{stiele10-11}, PE \citep{peacock10-10}, ZH \citep{zhang09-11}, VO \citep{voss06-07}, HO \citep{hofmann07-13}.}
\end{deluxetable*}

\begin{deluxetable*}{ c c c  c}
\tabletypesize{\scriptsize}
\tablecaption{Matching Analysis Summary \label{tab:match-results}}
\tablecolumns{4}
\tablewidth{0pt}
\tablehead{
\colhead{Type}		&
\colhead{Completeness Limit}		&
\colhead{Matched}	&
\colhead{Matched X-ray Sources}
}
\startdata
PHAT Star Clusters	&	M$_{F475W}=-4.0$ \& M$_{F814W}=-5.0$	&	15	&	17
\\	\htwo\ Regions		&	$L_{\rm{H}\alpha} = 10^{34}$ \es\	&	10	&	9
\\
\enddata
\tablecomments{The \htr\ catalogue \citep{azimlu10-11} reports a limiting luminosity but not a completeness limit; the star cluster catalogue \citepalias{johnson06-12} is 80\% complete to the indicated magnitudes. X-ray sources have a limiting luminosity of $8.5\times10^{33}$ \es.}
\end{deluxetable*}

\subsection{X-ray Properties of Matched Star Clusters \& \htwo\ Regions} \label{sec:xrmatches}

\begin{figure}[!h]
\plotone{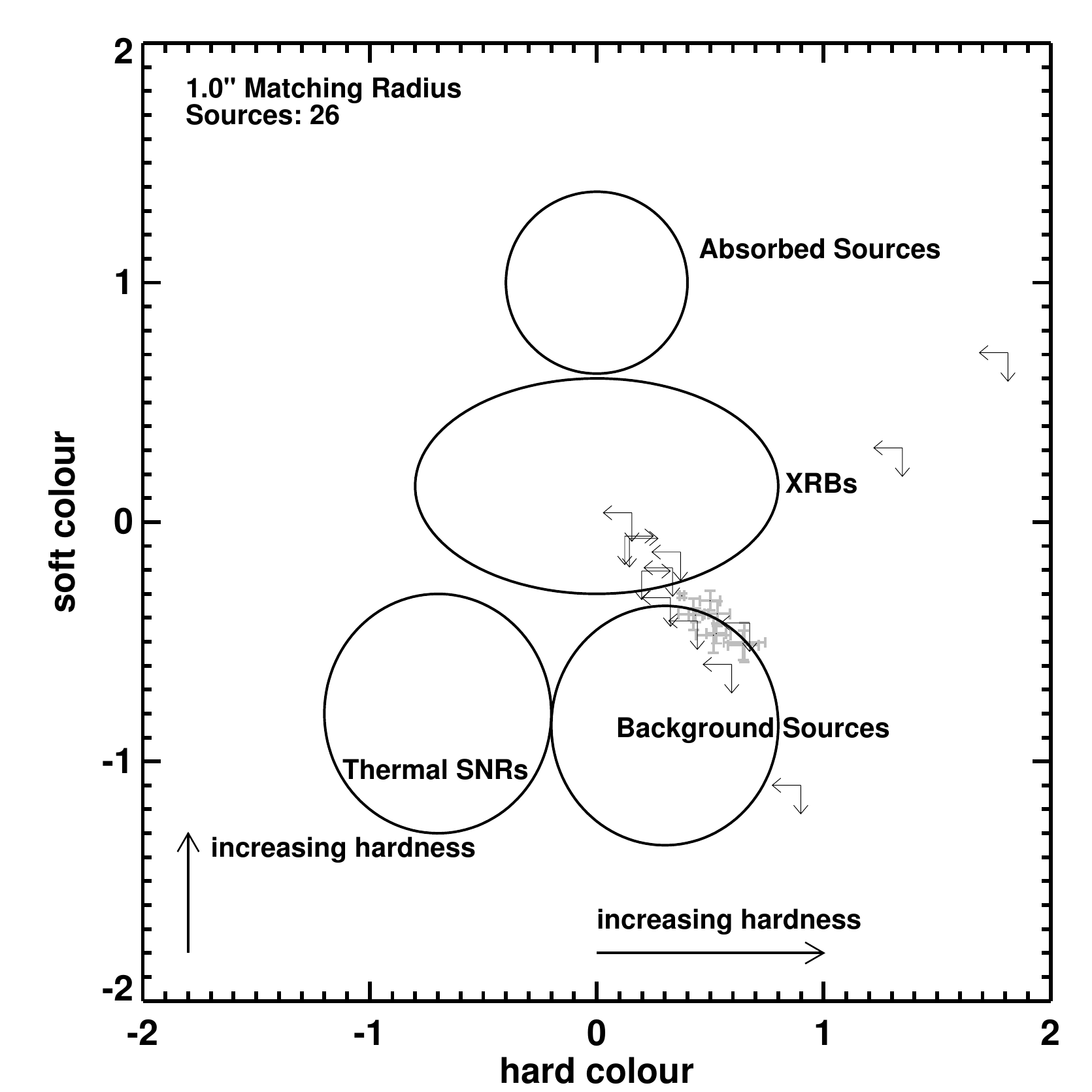}
\caption{X-ray color-color diagram of the 26 X-ray point sources in M31 that were matched to star clusters or \htwo\ regions. The X-ray color classification scheme of \citet{kilgard08-05} is overlaid. Sources that were not detected in one or more energy bands ($<3\sigma$) are plotted as upper or lower limits in one or both axes. Misclassification can result from the variation of XRBs spectral states over time. Also, since counts do not incorporate the change in effective area of \Chandra\ over time, additional uncertainty is introduced because observations are spread over 13 years. Most sources are found in the region of color space between XRBs and background sources, where previous classifications from Table \ref{tab:scxrmatches} confirm the majority of those matched to star clusters as GC-LMXBs.} 
\label{fig:xraycc}
\end{figure}

An X-ray color-color plot of the X-ray point sources matched to star clusters and \htwo\ regions is shown in Figure \ref{fig:xraycc}, with the X-ray color classification scheme of \citet{kilgard08-05} overlaid.
Soft and hard colors are defined by the soft (S, $0.3-1$ keV), medium (M, $1-2$ keV), hard (H, $2-8$ keV), and total (T, $0.3-8$ keV) band energies as $(M-S)/T$ and $(H-M)/T$ respectively. X-ray photometry was completed separately for each observation for each X-ray point source. The photometry results for each observation were then merged for each X-ray point source in each of the energy bands to determine colors. Due to the variation in the spectral states of XRBs, the same source can end up in different areas of the diagram. Also, only flux measurements account for the changing effective area of \Chandra, whereas counts do not. Since the observations we used were spread over 13 years, using counts introduces another source of uncertainty.
Absorption also moves X-ray sources into different regions of the diagram and thus can result in misclassifications. 
Most of the sources are clustered between the XRB and background source regions. However, given the low false match rate for star clusters and previous confirmations as GC sources it is likely that these are mostly XRBs. 
The X-ray source with AE ID \#3 has a luminosity typical of bright XRBs and was previously classified to match within 0.1\arcsec\ of an infrared source by \citet{kaaret10-02}. The remaining X-ray sources were already identified as GCs by various studies (see Table \ref{tab:scxrmatches} for refs.) with many that are likely black hole XRBs \citep{barnard06-13}.

\subsection{Optical Properties of Matched Star Clusters and \htwo\ Regions} \label{sec:opt-match}

\begin{figure}[!h]
\plotone{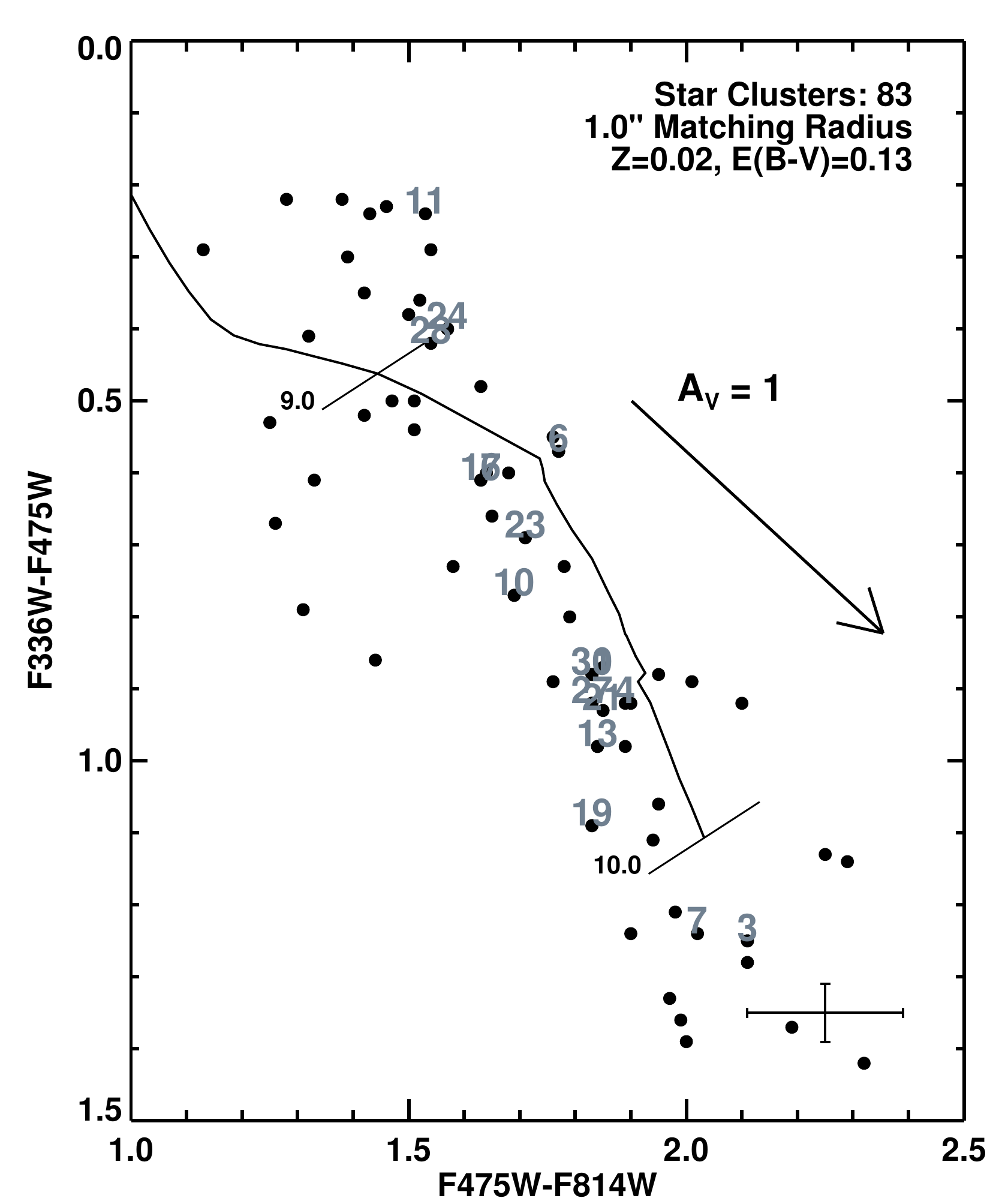}
\caption{Optical color-color diagram of the 83 star clusters in M31 from the catalogue of \citetalias{johnson06-12} that are within the field of view of the \Chandra\ X-ray image. A theoretical evolutionary track from the simple stellar population model of \citet{marigo05-08} with $Z = 0.02$ is included with 
a total reddening (external and internal) of $E(B-V) = 0.13$ applied \citep{caldwell02-11}. Bars on the track represent log(age) in years. Numbers labelling individual data points indicate the AE IDs of the 17 unique X-ray point sources that matched to 15 star clusters (Table \ref{tab:scxrmatches}). 
The foreground reddening vector is of length $A_{V} = 1$ mag.} 
\label{fig:sccc}
\end{figure}

A color-color diagram of the 83 star clusters \citepalias{johnson06-12} in the \Chandra\ coverage region is shown in Figure \ref{fig:sccc}. A theoretical evolutionary track from the simple stellar population models of \citet{marigo05-08} with $Z = 0.02$ is overlaid after applying a total reddening (external and internal) of $E(B-V) = 0.13$ \citep{caldwell02-11}. The 15 star clusters correspond to the 17 X-ray point sources they were matched to from Figure \ref{fig:xraycc} (results in Table \ref{tab:scxrmatches}). In Figure \ref{fig:sccc} most star clusters are $\geq$1 Gyr old as expected for an old bulge population.

\begin{figure}[!h]
\plotone{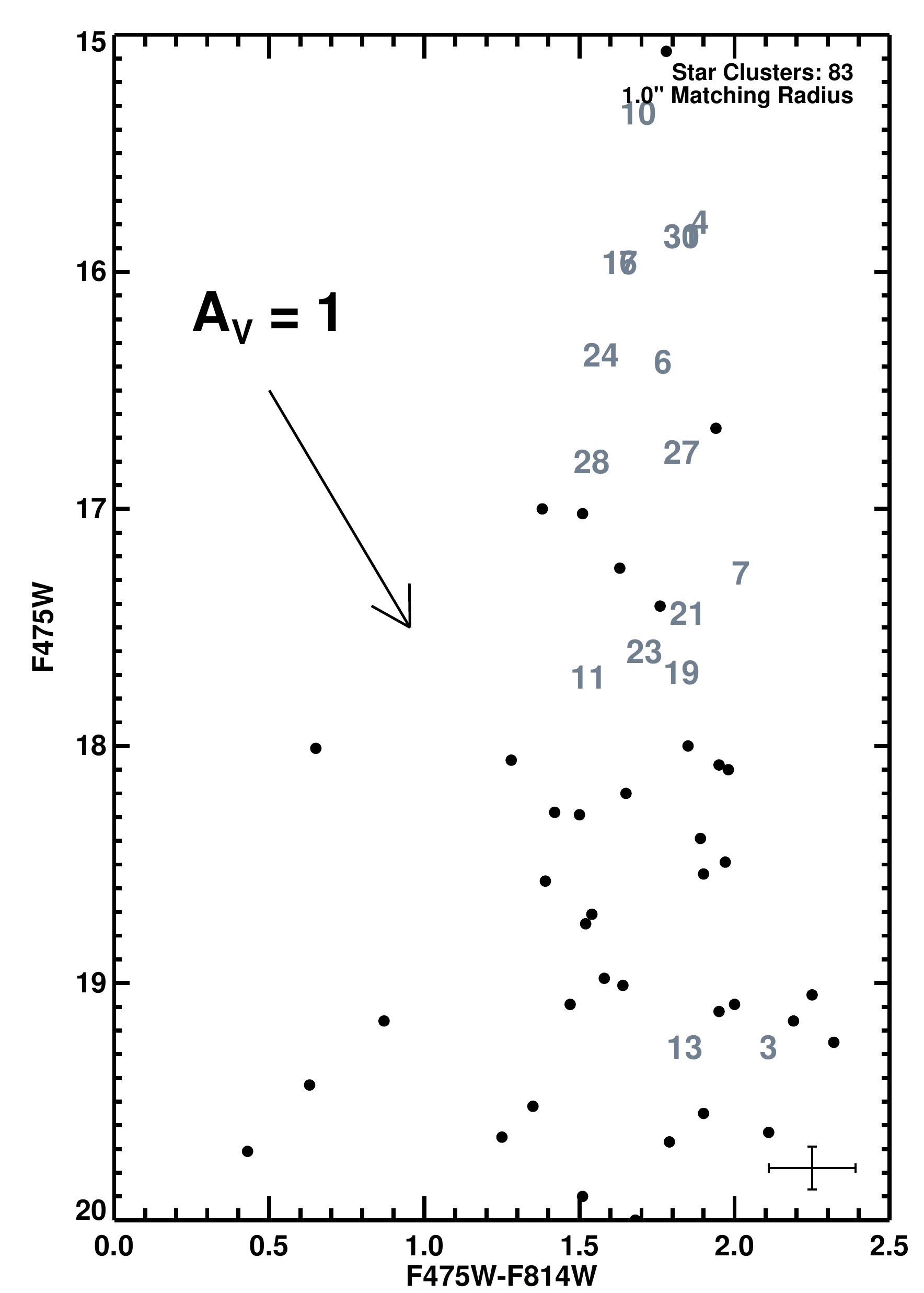}
\caption{Color-magnitude diagram of the 83 star clusters from the catalogue of \citetalias{johnson06-12} in the field of view of the \Chandra\ data in M31. Numbers labelling individual data points indicate the AE IDs of the 17 unique X-ray point sources that matched to 15 star clusters (Table \ref{tab:scxrmatches}). The foreground reddening vector is of length $A_{V} = 1$ mag. Most of the matched star clusters are brighter and redder than the remaining sample, as expected for clusters that host X-ray sources.}
\label{fig:cmd}
\end{figure}

\begin{figure*}[!ht]
\epsscale{0.50}
\begin{center}
\begin{tabular}{cc}
\plotone{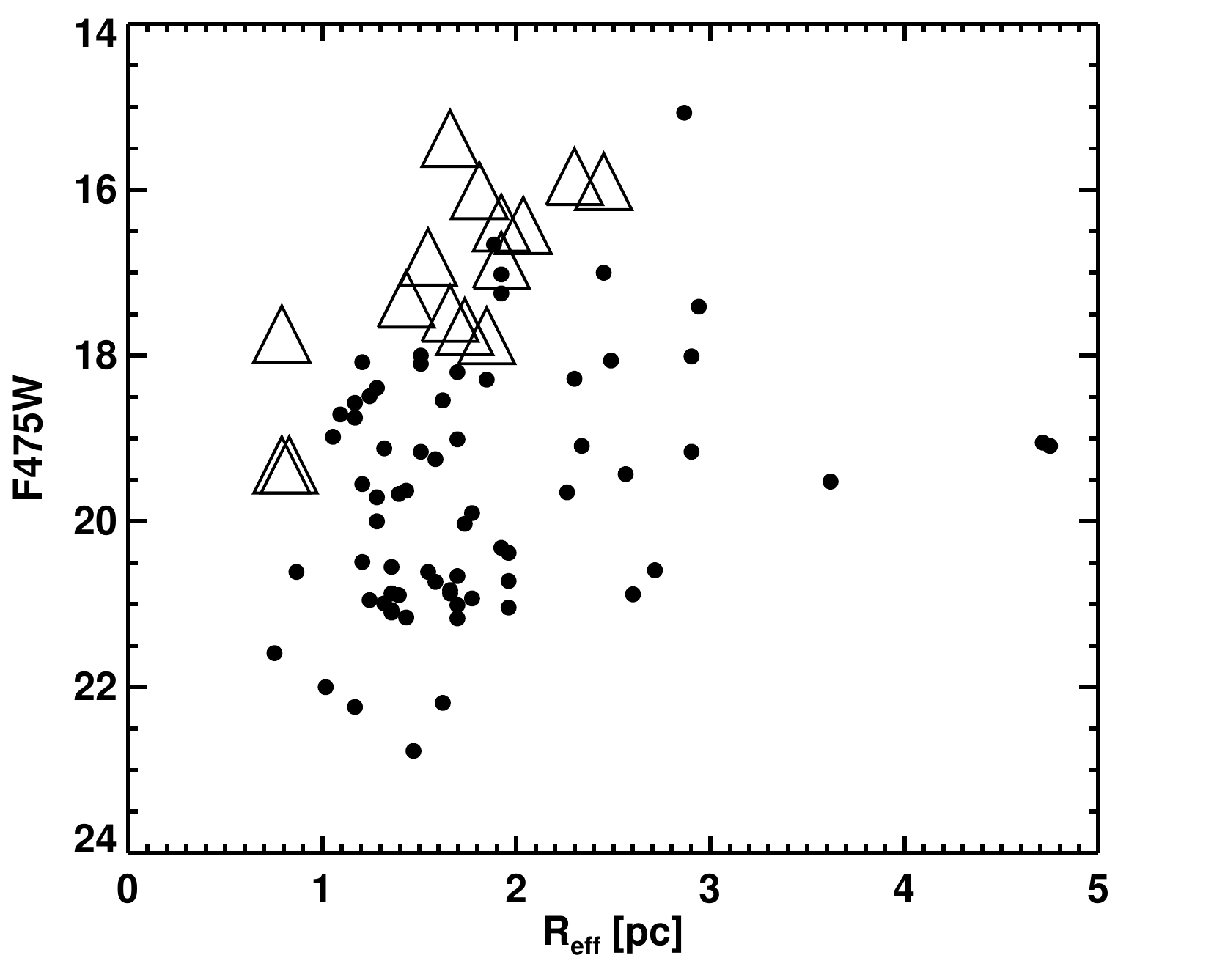}
\plotone{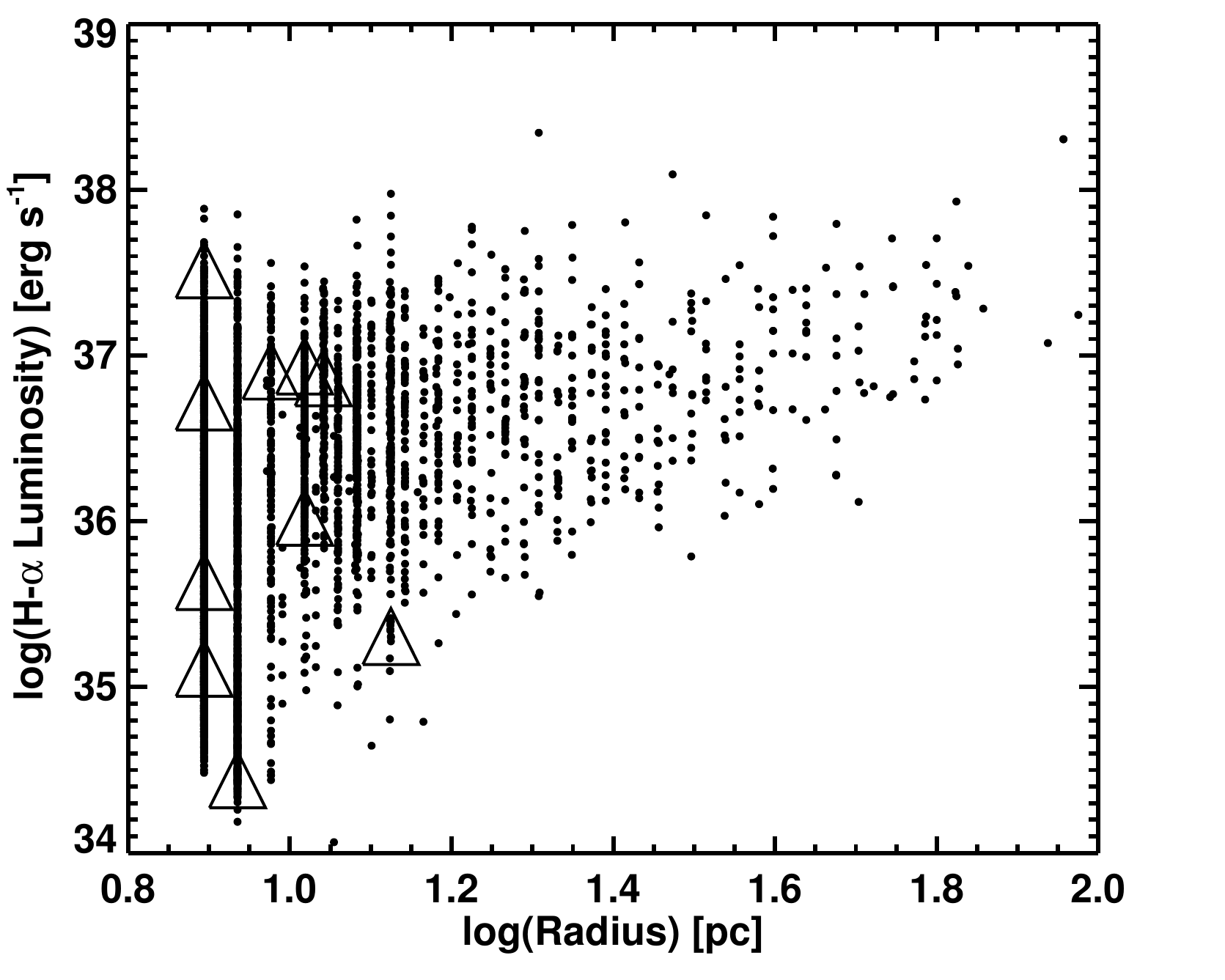}
\end{tabular}
\caption{F475W magnitude vs.\ effective radius $R_{\rm{eff}}$ for the 83 star clusters in M31 \citepalias{johnson06-12} that are within the field of view of the \Chandra\ data (left) and $L_{\rm{H}\alpha}$ vs. radius for the 1566 \htrs\ in the field of view of the \Chandra\ data (right). Triangles indicate the matches to X-ray point sources from Tables \ref{tab:scxrmatches} and \ref{tab:htwoxrmatches}. X-ray sources are preferentially found in luminous, compact star clusters. The apparent preference for compact \htrs\ to host X-ray sources is a artifact of their distribution (much larger number of compact \htrs). A statistical analysis (see Section \ref{sec:opt-match}) showed that neither the radius nor H$\alpha$ luminosity was a predictor of whether an \htr\ would host an X-ray source.} 
\label{fig:matches}
\end{center}
\end{figure*}

\begin{figure*}[!ht]
\epsscale{0.50}
\begin{center}
\begin{tabular}{cc}
\plotone{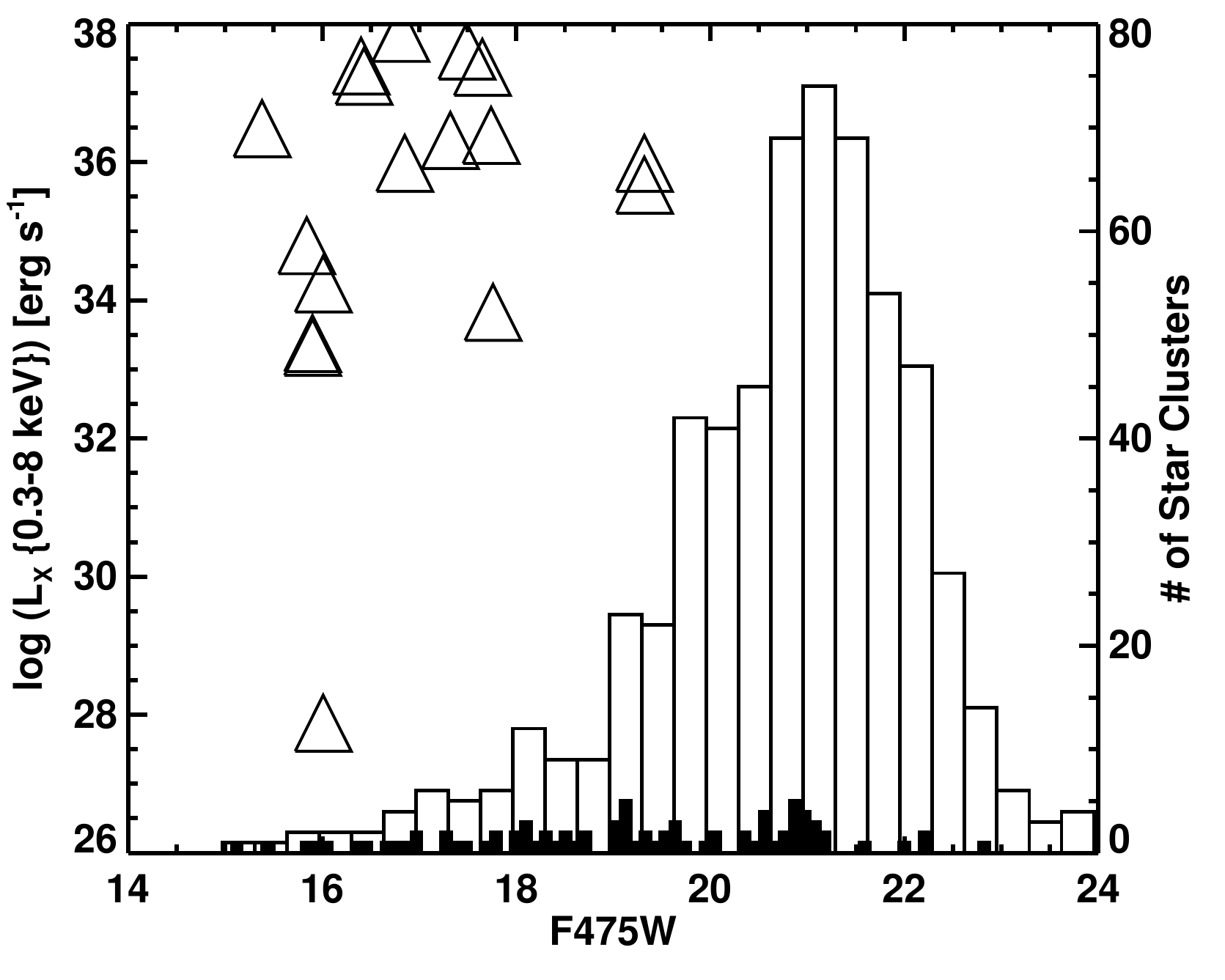}
\plotone{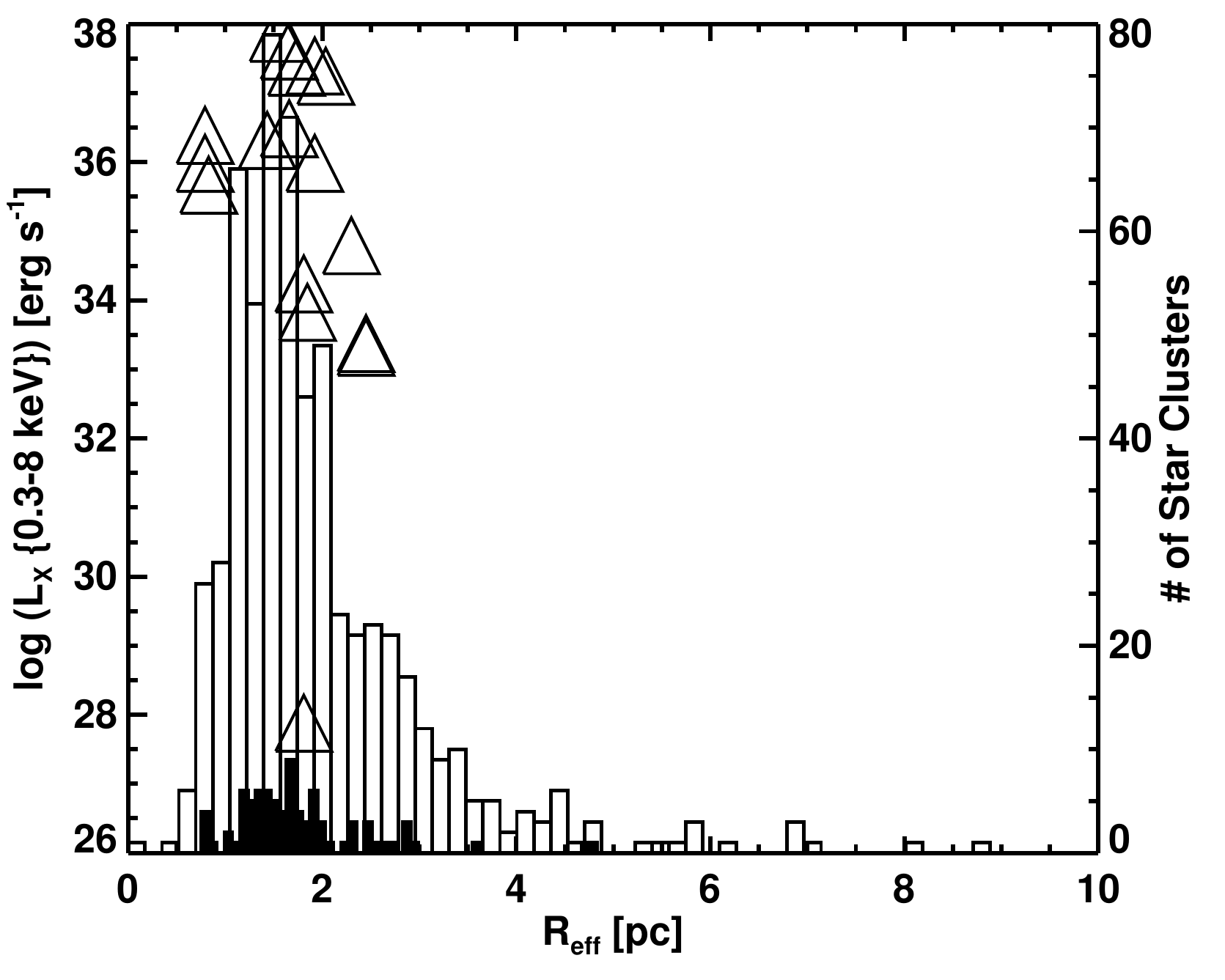}
\end{tabular}
\caption{The X-ray luminosity of star cluster matches compared with the F475W magnitude (left) and effective radius $R_{\rm{eff}}$ (right) of star clusters. The X-ray luminosity is that of the X-ray point source matched to the star cluster and was determined by ACIS Extract (see Section \ref{sec:xrdata-detect}). The 83 star clusters in the field of view of the \Chandra\ data are shown in the filled black histogram while the entire year 1 PHAT survey star cluster sample is represented by the larger unfilled histogram. Triangles indicate the matches to X-ray point sources from Tables \ref{tab:scxrmatches} and \ref{tab:htwoxrmatches}. Our sample has preferentially more bright clusters than the complete year 1 sample while the distributions of effective radius in the two samples appear consistent. A KS test confirmed these conclusions (see Section \ref{sec:opt-match}).}
\label{fig:histograms}
\end{center}
\end{figure*}

To determine whether our matches correspond to brighter and redder star clusters, we created the color-magnitude diagram shown in Figure \ref{fig:cmd}.  
It is evident that the matches are all brighter and redder than star clusters without a detected X-ray point source. This is in agreement with the studies of GC-LMXBs in both elliptical and spiral galaxies mentioned earlier \citep{sivakoff05-07, peacock10-10, paolillo08-11, mineo01-14}. The average F475W magnitude of matched star clusters was $17 \pm 0.6$ compared to $20.7 \pm 0.6$ for unmatched clusters, 
while the average effective radius of matched clusters was $1.6 \pm 0.1$ pc compared to $2 \pm 0.1$ pc for unmatched clusters\footnote{Uncertainties were not published for $R_{\rm{eff}}$ so we computed the standard error of the mean.}. This confirms the trend that more luminous, compact star clusters preferentially host an X-ray source. Further evidence is shown in the left panel of Figure \ref{fig:matches}, where it is clear that our matches are all above the 80\% completeness limit (m$_{F475W}=20.47$) mentioned in Table \ref{tab:match-results}. 
Figure \ref{fig:histograms} shows the X-ray luminosity of point sources that were matched to star clusters plotted against the F475W magnitude and effective radius $R_{\rm{eff}}$. The histograms show that our sample (dark portion) has preferentially more bright clusters than the complete year 1 sample (black outline - mostly disk clusters), while no significant variations in $R_{\rm{eff}}$ between the samples is seen. Specifically, the F475W magnitudes of our sample are skewed towards the bright end compared to the complete sample.
We performed a Kolmogorov-Smirnov (KS) test on the two distributions, namely our sample and the complete year 1 sample of star clusters, for both the F475W magnitude and $R_{\rm{eff}}$. We found that the probability that the two distributions are drawn from the same sample is $3.7\times10^{-14}$  for the F475W magnitude and 0.09 for $R_{\rm{eff}}$, confirming that the F475W magnitudes of the two distributions are different.

To assess the statistical significance of our results we used the $\bf{R}$ statistical software \citep{R-2008} to complete a logistic regression on our data. This allows us to assess the impact that a parameter has on the probability of a star cluster hosting an X-ray source. The generalized linear model ($glm$) in $\bf{R}$ is best-suited for this and has the form shown in equation \ref{eq:glm}, where $p_{i}$ is the probability of success averaged over $i$ trials (outcome of 1  or 0, i.e.\ matched or not) and the $\beta_{m}$'s are the regression coefficients of the variables $x_{m, i}$ that we model.
\begin{eqnarray}
ln\left(\frac{p_{i}}{1-p_{i}}\right) & = \beta_{0} + \beta_{1}x_{1,i} + \cdots + \beta_{m}x_{m,i}	\label{eq:glm}
\end{eqnarray}
For star clusters we modelled 3 variables, specifically the effective radius $R_{\rm{eff}}$, F475W magnitude, and color (F475W$-$F814W). We only included the 83 star clusters that were in the field of view of the \Chandra\ data, since only these star clusters could possibly have been matched to an identified X-ray point source. The results of the logistic regression are shown in Table \ref{tab:logit}. 

\begin{deluxetable}{ c c  c  c  c }
\tabletypesize{\scriptsize}
\tablecaption{Logistic Regression Results for Star Clusters and \htrs \label{tab:logit}}
\tablecolumns{5}
\tablewidth{0pt}
\tablehead{
\colhead{Variable}	&
\colhead{Coefficient $\beta$}	&		
\colhead{Standard Error}	&
\colhead{$z-$Value}	&
\colhead{Pr($>\abs{z}$)}
}
\startdata
\cutinhead{Star Clusters}
$R_{\rm{eff}}$	&   $-$20.7241	&     9.0953	&  $-$2.279	&  0.02269
\\	F475W	&    $-$2.9285	&     1.0582	&  $-$2.767	&  0.00565
\\	F475W $-$ F814W	&         0.5696	&    2.2257	&   0.256	&  0.79802
\\	\cutinhead{\htrs}
	 Radius	&     -0.9283	&      2.7518	&  -0.337	&    0.736
\\	H$\alpha$ Luminosity	&   13.9974	&     36.3963	&   0.385	&    0.701
\\
\enddata
\tablecomments{Results for a model fit to equation \ref{eq:glm}. The $z$-statistic in column 4 tells us how many standard error units the sample mean is from the population mean while the $p$-value in the last column gives us the probability that the null hypothesis is true. A variable in the model is significant for a $p$-value Pr($>\abs{z}$) $<$ 0.05.} 
\end{deluxetable}

Since a variable in the model is generally significant for a $p$-value Pr($>\abs{z}$) $<$ 0.05 (i.e.\ null hypothesis is rejected), only $R_{\rm{eff}}$ and the F475W magnitude are statistically significant. From Table \ref{tab:logit} the coefficient gives the $ln$ odds increase of a match for a unit increase in the respective variable. Therefore, a unit increase in $R_{\rm{eff}}$ results in a decrease of $\approx21$ in the $ln$ odds. However, because each variable has a different range of values, a unit increase does not have a 1:1 correspondence across variables. Therefore the smallest $p$-value indicates which variable is most important to the outcome. From this we see that the F475W magnitude is more significant when determining if a star cluster hosts a bright X-ray source, followed by the effective radius. However, star cluster color was not important, at odds with the surveys of GC-LMXBs in spiral and elliptical galaxies mentioned above. These surveys studied entire galaxy populations, meaning that even if GC catalogues were incomplete their numbers were not radically biased towards metal-poor or metal-rich. Our result could be a consequence of our sample since it was restricted mostly to bulge clusters, which are metal-rich (see Figure \ref{fig:cmd}). Indeed, a KS test between the F475W$-$F814W color of star clusters in our sample and the complete year 1 PHAT clusters gives a probability of $1.7\times10^{-13}$ that the two distributions are drawn from the same sample. This confirms that bulge clusters are more metal-rich than those in the disk. Complete surveys of M31's GCs by \citet{peacock10-10} and \citet{agar11-13} showed that metal-rich star clusters were more likely to host an X-ray source, indicating our result is biased by our sample.

The 10 \htwo\ regions matched to X-ray sources have average radii and H$\alpha$ luminosities of $R=9.4 \pm 0.6$ pc and $L_{\rm{H}\alpha} = 1.4 \pm 0.2 \times10^{36}$ \es\ compared to averages of $R=11.2 \pm 0.2$ pc and $L_{\rm{H}\alpha} = 1.2 \pm 3.2 \times10^{36}$ \es\ for unmatched \htwo\ regions\footnote{Uncertainties were not published for radii and H$\alpha$ luminosities so we computed the standard error of the mean.}. While \htrs\ that were matched to X-ray point sources seem to favour those that are more compact, we cannot claim a trend exists with H$\alpha$ luminosity due to the large errors associated with their average values. The right panel of Figure \ref{fig:matches} shows the distribution of \htr\ luminosity and radius, with matches crowding the compact end. However, \htwo\ regions have a much wider range of radii than globular clusters and generally are not well approximated by circular apertures \citep{azimlu10-11}. These differences complicate comparisons between the populations. The limiting $L_{\rm{H}\alpha}$ for \htrs\ from Table \ref{tab:match-results} applies across the entire $R_{\rm{eff}}$ range; those with larger radii naturally have a larger luminosity and are not biased by incompleteness.

As for star clusters we also completed logistic regression on the 1566 \htrs\ that were in the field of view of the \Chandra\ data. We modelled the radius and H$\alpha$ luminosity of \htrs\ to assess their impact on the probability of hosting an X-ray source, with the results shown in Table \ref{tab:logit}. 
From the $p$-values of both variables we can see that neither is significant in determining the presence of an X-ray source as both are $\gg0.05$. 
In the right panel of Figure \ref{fig:matches} it would seem that at least the most compact \htrs\ were more likely to host an X-ray source. However, based on the distribution of the population, specifically that most \htrs\ have small radii, this impression is not supported by our statistical analysis.

\section{X-Ray Stacking} \label{sec:stacking}

To study the faint population of X-ray sources we performed a stacking analysis of the 68 unmatched star clusters and 1556 unmatched \htwo\ regions in M31. Using the techniques of \citet{brandt07-01, brandt09-01} and \citet{hornschemeier06-01} we stacked 31 by 31 (45 by 45) pixel regions centered on star cluster (\htwo\ region) positions in X-ray images of M31.

\begin{figure*}[!ht]
\begin{center}
\epsscale{.5}
\begin{tabular}{cc}
\plotone{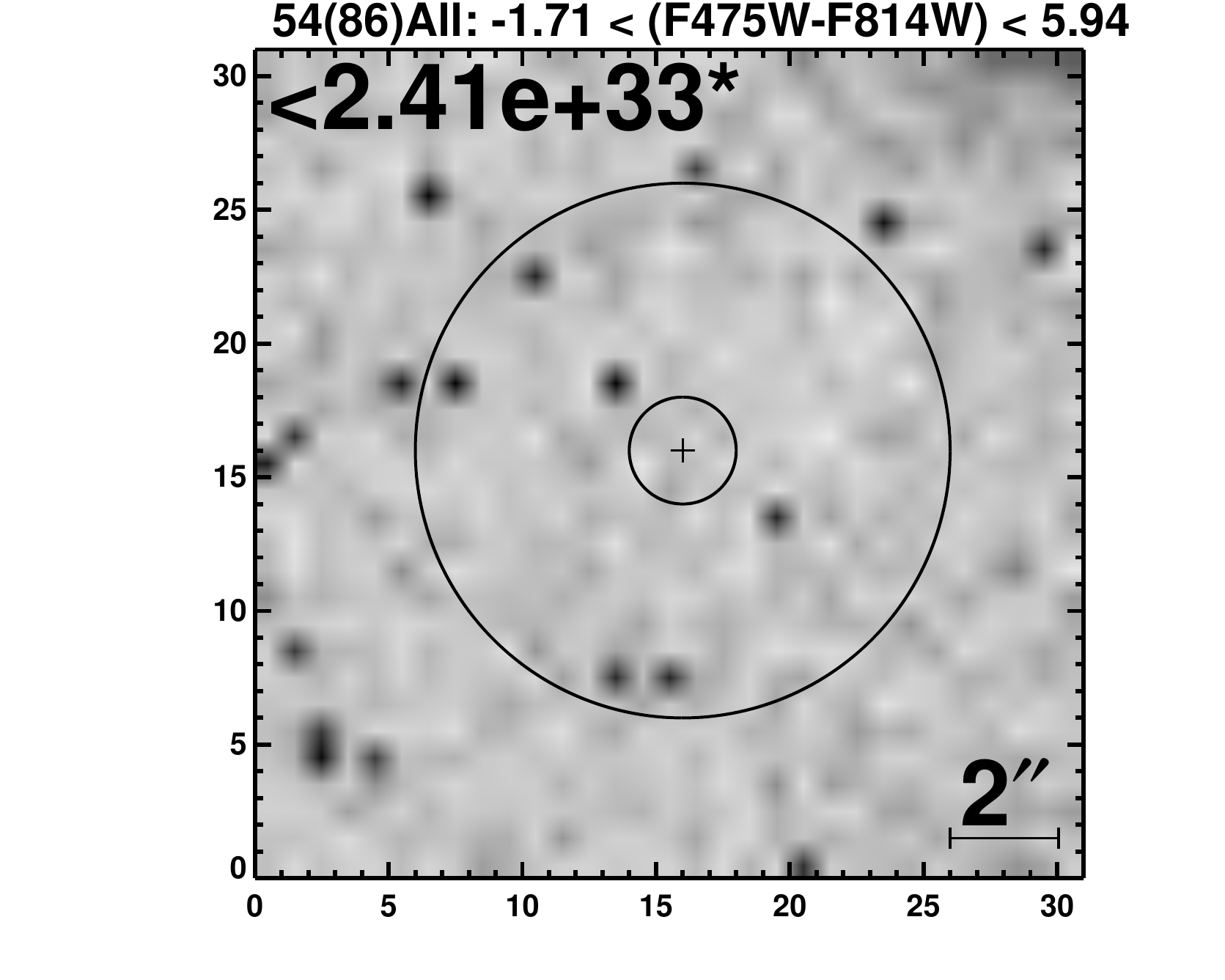}
\plotone{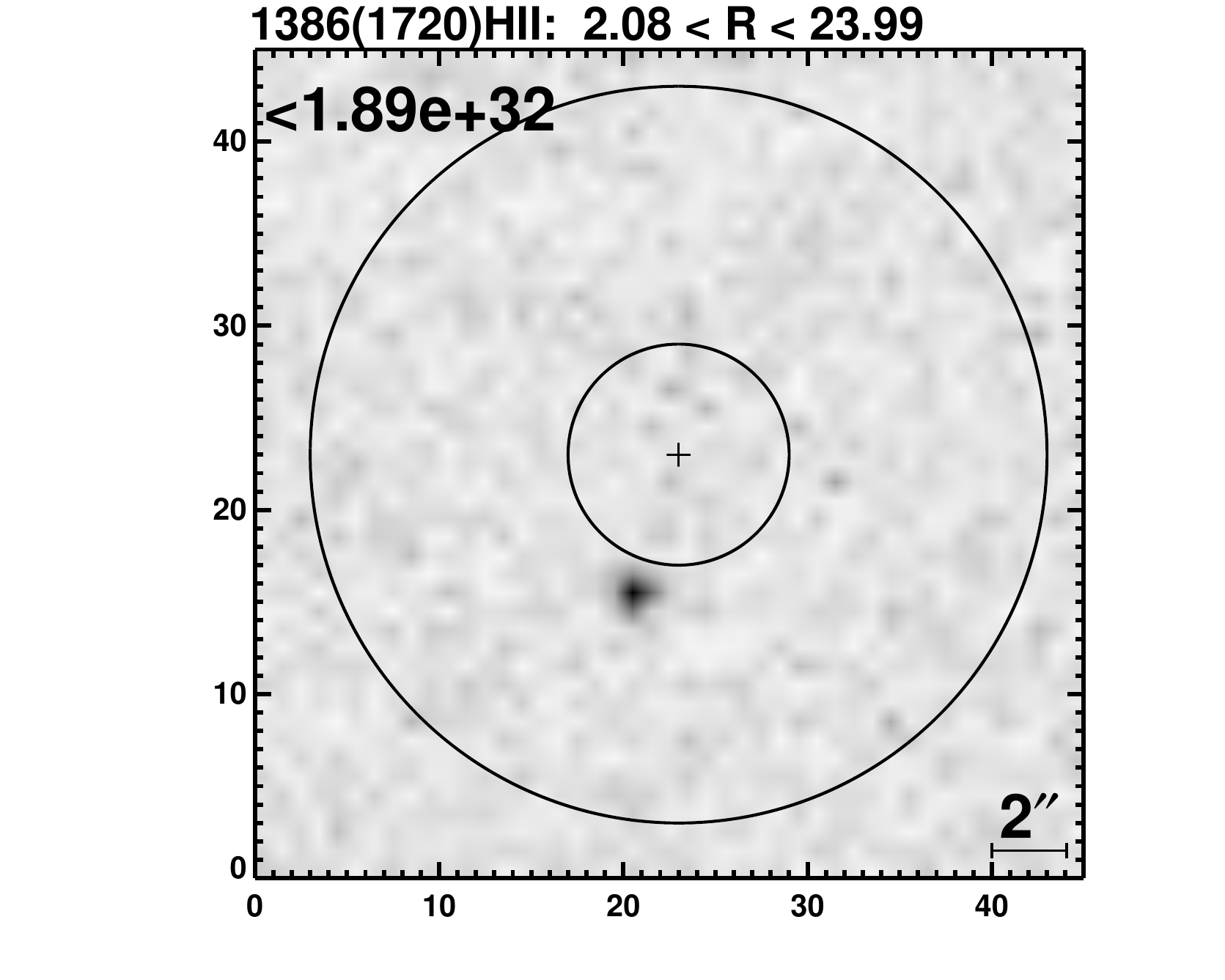}
\end{tabular}
\caption{{Stacked \Chandra~images ($0.3-8$ keV) of all 54 unique star clusters (left) and 1386 unique \htrs\ (right) in M31 (total number from ACIS-I and ACIS-S in brackets). Star cluster 
color range (\htr\ radius in pc) is also indicated. 
Pixels are 0.492$\arcsec$ with 31 (45) per side. North is up and east is to the left. The source aperture is 4 (12) pixels in diameter while the background was calculated outside the larger aperture of diameter 10 (20) pixels. The value in the top left corner indicates the 1.3$\sigma$ upper limit per star cluster (\htr) in the stack on the source luminosity in \es\ (asterisk denotes a negative luminosity in the source aperture). The grayscale represents the brightest pixel (darkest) relative to the dimmest pixel (lightest). All stacked images of star clusters and \htrs\ across all energy bands resulted in non-detections.}} 
\label{fig:stack}
\end{center}
\end{figure*}

Star cluster (\htwo\ region) positions closer than 16 (23) pixels to the edge of an X-ray image were excluded since each image in the stack has to be complete (i.e.\ cannot have null data in any pixels). This left a total of 54 star clusters and 1386 \htwo\ regions in each sample. Since there are only a handful of star clusters and they are all older than 1 Gyr we do not subdivide the population by color. However, because brighter and smaller clusters preferentially host X-ray sources we do filter\footnote{We chose our filter cutoffs based on the histograms for radius and magnitude.} by $R_{\rm{eff}}$ and F475W magnitude (see Table \ref{tab:stackvals}) to take advantage of these trends. We followed a similar method for \htrs. There are two peaks in the luminosity function for \htrs\ in M31, at 10$^{35}$ \es\ and $4\times10^{36}$ \es\ for B stars and O stars respectively \citep{azimlu10-11}. These values were used to separate the population by luminosity when stacking. We stacked star clusters and \htrs\ in the full ($0.3-8$ keV), hard ($2-8$ keV), and soft ($0.3-2$ keV) energy bands. 
Sample stacked images of star clusters and \htrs\ are shown in Figure \ref{fig:stack}.

\begin{figure*}[!ht]
\epsscale{0.50}
\begin{center}
\begin{tabular}{cc}
\plotone{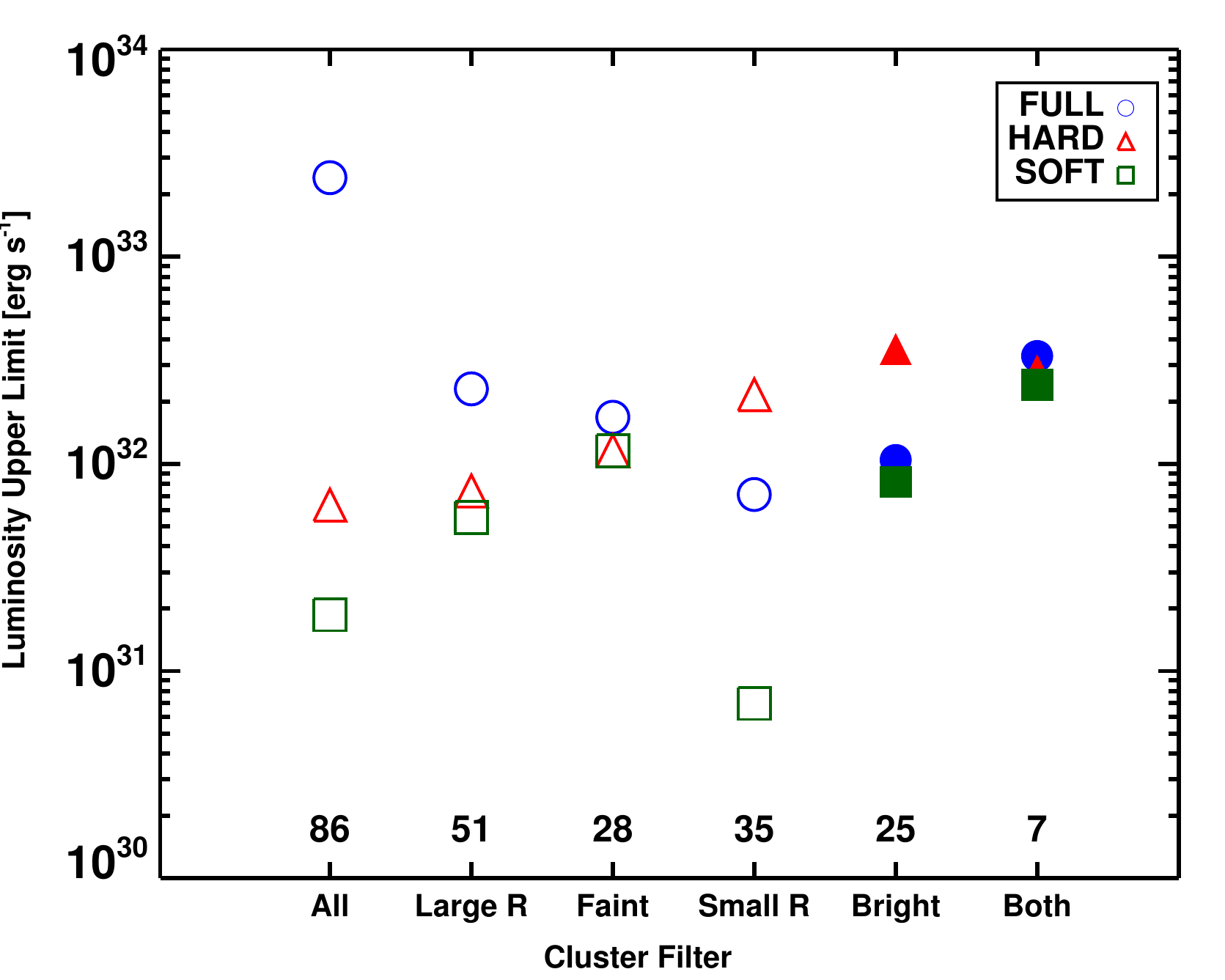}
\plotone{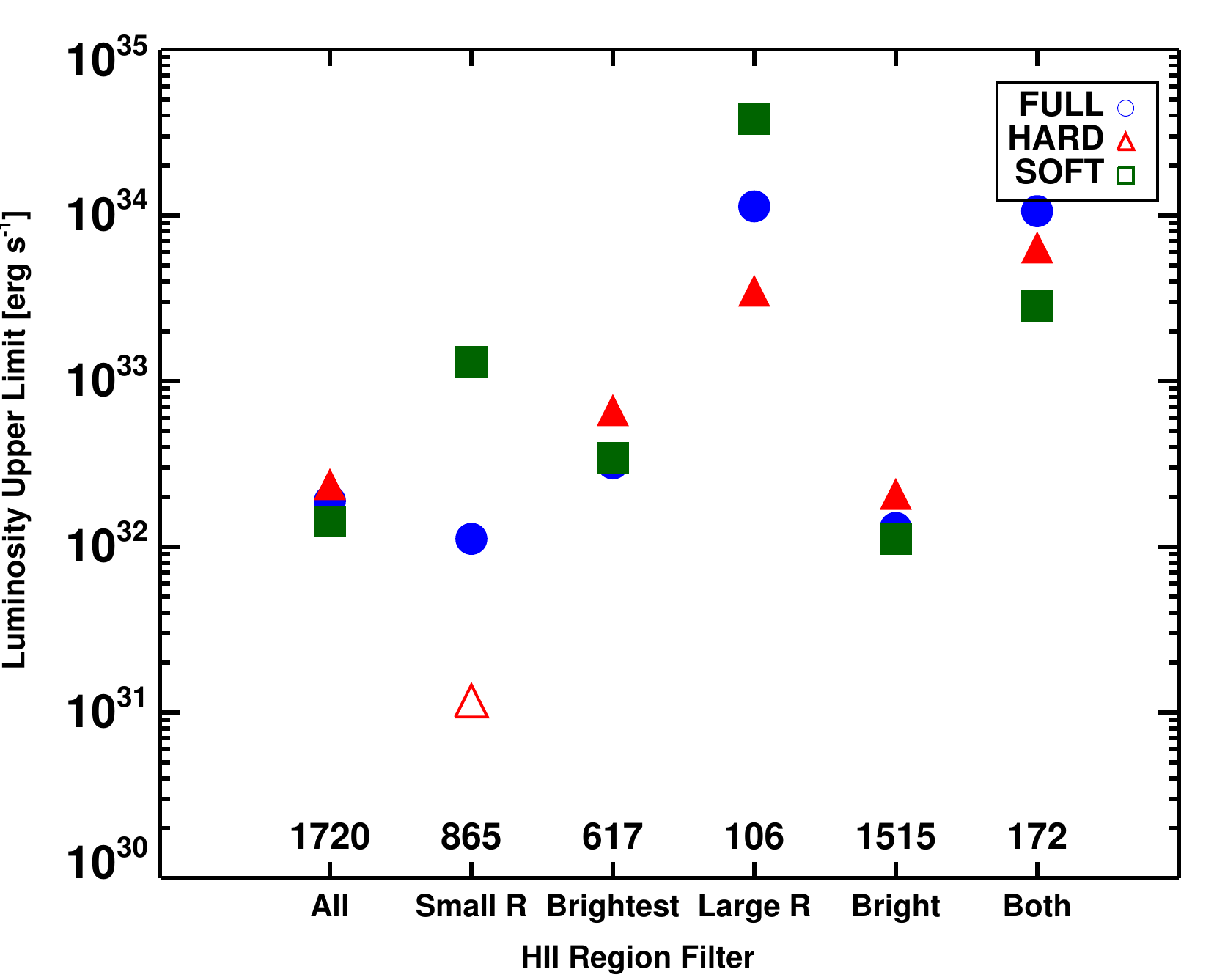}
\end{tabular}
\caption{Upper limits in \es\ for a specific filter type in the full band (circles), hard band (triangles), and soft band (squares) for star clusters (left) and \htrs\ (right). Values above the x-axis indicate the total number of stacked objects (see Table \ref{tab:stackvals} for the filter ranges). Filled shapes represent positive source luminosities while empty shapes are negative source luminosities (background-subtracted source aperture counts $<$0). When stacking star clusters with the `bright' and `both' (bright and small) filters we see that the source aperture luminosity becomes positive (average pixel value is larger than the background). This can be attributed to the higher probability of brighter and (in combination) more compact clusters hosting X-ray sources. Star cluster upper limits are approximately two orders of magnitude fainter than those found for M51 \citep{vulic02-13}.} 
\label{fig:limits}
\end{center}
\end{figure*}

The net counts in the source aperture $C_{n} = C_{s} - C_{b}\times A_{s}/A_{b}$, where $C$ and $A$ represent the counts and area respectively for the source (s) and background (b) regions. The source region for star clusters in a stacked image is 1\arcsec\ (2 pixels) in radius, which corresponds to the radius used to match star clusters to X-ray point sources. Keeping the source aperture radius consistent with the matching radius ensures that we do not include any X-ray sources outside of the original matching radius in our stacked images. The background region is defined to be outside the 90\% encircled energy radius of the source aperture, which for star clusters that are 4\arcmin\ off-axis from the \Chandra\ aimpoint is 10 pixels\footnote{Most star clusters are found in the bulge and thus are within $\sim$4\arcmin\ of the aimpoint for ACIS observations (centered on the supermassive black hole).}. \htrs\ are more extended and also randomly distributed throughout the ACIS field of view compared to star clusters. Average off-axis angles for \htrs\ are not much larger than 8\arcmin\ and so we use a 20 pixel diameter for the background region (corresponding to 90\% encircled energy radius for sources 8\arcmin\ off-axis). A negative source aperture luminosity results when the background-subtracted source aperture counts are $<$0. This means that the background has a larger average value per pixel than the source aperture. We use Poisson statistics (method of \citealt{gehrels04-86}) to calculate our uncertainties. We computed the 3$\sigma$-clipped mean $\mu$ ($C_{b}/A_{b}$) in the background, where pixels with counts $>\mu$ + 3$\sigma$ were excluded. 
Upper and lower limits for both the source and background regions were calculated using equations (9) and (14) from \citet{gehrels04-86}, using $S = 1.282$ (number of Gaussian $\sigma$) and confidence level parameters $\beta = 0.01$ and  $\gamma = -4.0$. 
Uncertainties in the net counts C$_{n}$ were calculated by normalizing background values to those of the source aperture. 
Luminosities for stacked images were computed from fluxes\footnote{See section \ref{sec:xraydata} for details on exposure-corrected images.} (photons cm$^{-2}$ s$^{-1}$) using a power-law spectrum with $\Gamma$=1.7 for each energy band and assuming a distance of 776 kpc. 
Uncertainties for luminosities were determined by multiplying the ratio of luminosity to counts in the source aperture by the uncertainties in counts. 
We use a larger size for our stacked images of \htrs\ (45 pixels) than star clusters (31 pixels) since \htrs\ have larger average radii. 
A caveat to our method is that our merged X-ray images of M31 have an inconsistent PSF at any point in the complete ACIS-I or ACIS-S image. This is a result of observations overlapping at different off-axis angles. This compounds the problem of stacking since we already introduce the same type of effect by stacking objects at various off-axis angles. All stacking completed on star clusters and \htrs\ resulted in non-detections. The results are displayed in Table \ref{tab:stackvals} and shown graphically in Figure \ref{fig:limits}. Correcting for foreground absorption (assuming a Galactic $N_{\rm H}$ value of $6.6\times10^{20}$ cm$^{-2}$ \citep{dickey-90} and a power-law spectrum of $\Gamma\sim1.7$) would increase all our luminosities by $\sim14\%$.

\section{Discussion}	\label{sec:discuss}

\subsection{Star Clusters}

Since we analyzed the bulge of M31, which is comprised of an old stellar population and thus globular clusters, it is not surprising that our X-ray point sources are predominantly GC-LMXBs (Table \ref{tab:scxrmatches}). When comparing the averages of star cluster colour, magnitude, and effective radius our results showed that redder, brighter, and compact star clusters were more likely to host X-ray sources. However, a logistic regression analysis showed that the F475W magnitude is the most significant predictor followed by the effective radius. The F475W$-$F814W color (an indicator of metallicity in old stellar populations) was not a significant predictor of whether a cluster hosted an X-ray source. A statistical analysis using logistic regression is a more robust method than comparing averages and as such accurately represents trends in the data. \citet{peacock10-10} found different trends when using \xmmn\ observations of M31 to identify 45 GC-LMXBs, namely that the stellar collision rate (proportional to stellar density and core radius), luminosity, and metallicity are all significant. However, \citet{agar11-13} came to similar conclusions to ours when using \HSTt\ data to derive structural parameters for 29 GC-LMXBs, and complementing the sample with published values for a total of 41 GC-LMXBs and 65 non-LMXB GCs. They found the probability of a GC hosting an X-ray source increased with increasing collision rate and proximity to the galaxy center. Metallicity was not as important a predictor but an increasing cluster mass at fixed collision rate decreased the probability of hosting an X-ray source (although the authors stress this may be a sample selection effect). The latter result does not agree with our positive trend between GC magnitude (which scales with mass) and a GC's probability of hosting an LMXB. This discrepancy could be due to our limited sample of bulge GCs that do not have as wide a range in metallicity as GCs at larger galactocentric distances. 

Studies of the GC-LMXB connection in spiral galaxies are limited. \citet{rangelov10-12} studied XRBs in the Antennae galaxies and found that massive young star clusters are more likely to host an X-ray source than less massive young clusters. When studying 32 GC-LMXBs in the spiral galaxy M104 with \Chandra, \citet{di-stefano12-03} found that metal-rich GCs were more likely to host an X-ray source. \citet{bellazzini02-95} found similar trends in the Milky Way. \citet{peacock10-10} does mention that the observed trends in M31 are weaker than those found for elliptical galaxies, which further stresses the need to analyze the updated star cluster data from PHAT with \Chandra, along with surveys of other spiral galaxies.

Our stacking results gave upper limits of $\approx10^{32}$ \es\ across all bands for different stacks. When stacking star clusters that were brightest (F475W magnitude), most compact (smallest $R_{\rm{eff}}$), and a combination of the two, we found that the detection significance\footnote{aperture luminosity / aperture upper limit} increased across all bands, although not to a statistically significant level. 
Due to the proximity of M31 the upper limits are approximately two orders of magnitude smaller than for star clusters in M51 \citep{vulic02-13}.  Studies of globular clusters in the Milky Way have found a plethora of faint X-ray sources that include quiescent XRBs (typically $L_{X} \sim 10^{33}$ \es) \citep{heinke06-05, heinke11-06, heinke02-09, heinke05-10, bahramian01-14}. In GC 47 Tucanae, five classified quiescent LMXBs had a mean $L_{X} (0.5-2.5$ keV) $\approx10^{32}$ \es\ \citep{heinke06-05}. Therefore, assuming M31's GCs are similar to those in the Milky Way, we should be probing this population with our stacking analysis. Given that a single GC may only have a handful of faint LMXBs, even our stacked sample of 4 of the brightest and smallest GCs fails to provide a detection. 
At this point it seems that stacking star clusters to identify faint XRBs is proving difficult, at least in spiral galaxies/crowded fields. With the higher concentration of X-ray sources in galaxy bulges source crowding is a problem. Along with diffuse emission in spiral disks, this complicates stacking analyses. Specifically, the background level of each image in a stack is larger making it more difficult to detect faint sources in the aperture. 
Ultimately a detailed, deep survey of a large sample of GCs in an early-type galaxy would be most promising for obtaining a stacking detection.

\subsection{\htwo\ Regions}

\htrs\ emit thermal X-rays at $kT \sim 0.2-0.3$ and $0.6-0.7$ keV due to shocks from OB star winds and supernovae/supernova remnants \citep{strickland12-00, grimes07-05, tullmann12-09, mineo11-12, li01-13}. The lifetimes of \htrs\ (a few Myr) are long enough that HMXBs can form, providing another possible source of X-ray emission. Therefore we would expect X-ray emission in \htrs\ to come from various sources, with bright HMXBs and supernova remnants (if present) dominating. 
\citet{tullmann12-09} found an HMXB in the giant \htr\ IC131 in M33, while \citet{ducci05-13} found 2 HMXB candidates in \htrs\ in M83. \citet{berghea10-13} studied ultraluminous X-ray sources in nearby galaxies and concluded that 18 of 27 were located inside OB associations or star forming regions. 
However, most HMXBs are not found in \htrs\ or OB associations. In the Milky Way, \citet{bodaghee01-12} found clustering between 79 HMXBs and 458 OB associations (7$\sigma$ for distances $<1$ kpc). The average offset of 0.4 $\pm$ 0.2 kpc between HMXBs and OB associations was consistent with compact object natal kicks of 100 $\pm$ 50 km s$^{-1}$. The average kinematical age (time from supernova to HMXB phase) was $\sim$4 Myr. A follow up to this work by \citet{coleiro02-132} found the upper limit on kinematical age to be $\sim$3 Myr. Therefore unless natal kicks are small HMXBs could be located $\approx$ 6.5\arcmin\ (400 pc) from \htrs\ in M31. Based on these results one would expect most X-ray emission from \htrs\ to be thermal rather than from XRBs. 

From Table \ref{tab:htwoxrmatches} most of the previous classifications for \htrs\ hosting X-ray point sources in M31 have been supernova remnants as opposed to HMXBs. However, these are only supernova remnant candidates that have been found using optical line ratios (H$\alpha$/S~{\sc ii}). In addition, our X-ray color-color plot (Figure \ref{fig:xraycc}) shows that none of the X-ray point sources appear in the supernova remnant regime. When comparing the average radii of matched and unmatched \htrs\ those that were more compact were the most likely hosts of X-ray sources. 
Nevertheless, a logistic regression analysis showed that both the luminosity and radius of an \htr\ were not significant predictors of it hosting an X-ray source. The four \htr\ matches that are previously unidentified can be classified as HMXB candidates and require further analysis to determine their nature. 
As with star clusters our stacking results of \htrs\ gave upper limits of $\approx10^{32}$ \es\ across all bands for most stacks. We saw the most improvement in detection significance (soft band) when filtering by the smallest radii ($<8.5$ pc) and $L_{\rm{H}\alpha} > 4\times10^{36}$ \es, although not to a statistically significant level.

\subsection{\Chandra\ Coverage of M31}
While most of the \Chandra\ observations have been restricted to the central region of M31, an extension of the data to the disk is necessary to probe both young star clusters and the numerous \htrs. The resolution of X-ray data that does cover most of M31 (i.e.\ \xmmn\ and \rosat) is not sufficient to distinguish individual X-ray point sources in crowded regions or reliably associate them with clusters. More specifically, with the exquisite resolution of \HSTt\ observations such as the PHAT survey only \Chandra\ will allow for robust identification of X-ray counterparts. 

Figure \ref{fig:acis-fov} is truly striking. While \Chandra\ ACIS-I data does cover a large portion of the first 8 bricks of the PHAT survey, the remaining bricks and almost the entire northeast quadrant of M31 are unobserved. In M31, \Chandra\ has been used mostly for a long-term monitoring program of the supermassive black hole in the nucleus. Other observations throughout the galaxy were used to study supersoft X-ray sources. Due to the degradation of the PSF at large off-axis angles, data outside of the ACIS-S3 chip has poor resolution and therefore is not useful even for stacking. Further motivation lies in the absence of a confirmed HMXB in M31, which can be addressed by using both the \htr\ and star cluster catalogues available to identify candidates. Investigating the spiral arm regions that the PHAT survey has observed with (future) complementary \Chandra\ data would be invaluable to furthering our understanding of the XRB population in M31.

\section{Summary}	\label{sec:summary}

We used all 121 publicly available \Chandra\ ACIS observations of M31 to complete a study of star clusters and \htrs. Specifically, we analyzed 83 star clusters in the bulge of M31 from the PHAT survey and found 15 unique matches (18\%) to 17 X-ray point sources within 1\arcsec\ (3.8 pc). When correcting our percentage of matches to completeness limits of past surveys we found similar results. The most compact, brightest, and reddest clusters preferentially hosted X-ray sources, on average. Logistic regression showed that the star cluster F475W magnitude was the most important predictor of finding a star cluster X-ray source. This follows from the fact that the most luminous and therefore most massive star clusters are more likely to host XRBs. A less significant predictor was $R_{\rm{eff}}$, which associated smaller star clusters (probes compactness to some extent) with X-ray sources. The link between these two properties and XRBs stems from the high densities and larger number of stars that create dynamical conditions most favorable for the formation of XRBs. The F475W$-$F814W color, an indicator of metallicity in old stellar populations, was not a significant predictor for star clusters hosting X-ray point sources. The majority of clusters in our sample are metal-rich and thus we do not have an evenly distributed sample of blue and red clusters. A more complete survey with a larger spread in metallicity is required to confirm this result. 
A stacking analysis resulted in non-detections across all energy bands and average upper limits of $\approx10^{32}$ \es. This is consistent with quiescent XRB luminosities in the Milky Way. 

From 1566 \htrs\ in the field of view of the \Chandra\ data only 10 matched to 9 unique X-ray point sources within 3\arcsec\ (11 pc). On average, the matches corresponded to the most compact \htrs. However, a logistic regression analysis showed that neither the radius nor H$\alpha$ luminosity was a significant predictor of an \htr\ hosting an X-ray source. From previous optical emission-line classifications most of these sources were supernova remnant candidates. Four sources that have no previous optical identifications are HMXB candidates. A stacking analysis resulted in upper limits of $\approx10^{32}$ \es\ for all energy bands. To advance our understanding of the XRB population in M31, specifically the disk, and probe the faintest extragalactic XRBs, a complete \Chandra\ survey of the PHAT region is necessary.

\acknowledgements
We thank the referee for detailed comments that improved the manuscript. We also thank Ben Williams for helpful comments and Cliff Johnson for the PHAT footprint. Support for this work was provided by Discovery Grants from the Natural Sciences and Engineering Research Council of Canada and by Ontario Early Researcher Awards. We have also used the Canadian Advanced Network for Astronomical Research (CANFAR; \citealt{gaudet07-11}). This research has made use of the NASA/IPAC Extragalactic Database (NED), which is operated by the Jet Propulsion Laboratory, California Institute of Technology, under contract with the National Aeronautics and Space Administration. We acknowledge the following archives: the Hubble Legacy Archive (\url{hla.stsci.edu}), Chandra Data Archive (\url{cda.harvard.edu/chaser}), and 2MASS (\url{ipac.caltech.edu/2mass}). \\ 
\indent \emph{Facilities:} HST (ACS, WFC3), CXO (ACIS)

\bibliographystyle{aa}

\clearpage
\begin{turnpage}
\begin{deluxetable*}{c c c c c c c c c c c c}
\tabletypesize{\scriptsize}
\tablecaption{Stacked Image Properties \label{tab:stackvals}}
\tablecolumns{12}
\tablewidth{0pt}
\tablehead{
\colhead{Filter} &
\colhead{Unique Number} &
\colhead{Total Number\tablenotemark{a}} &
\multicolumn{3}{c|}{FULL} &
\multicolumn{3}{c|}{HARD} &
\multicolumn{3}{c}{SOFT} \\
\cline{4-6} \cline{7-9} \cline{10-12}
\colhead{} &
\colhead{} & 
\colhead{} &
\colhead{Upper Limit\tablenotemark{b}} &
\colhead{Sky\tablenotemark{c}} &
\colhead{Sky Sigma\tablenotemark{d}} &
\colhead{Upper Limit\tablenotemark{b}} &
\colhead{Sky\tablenotemark{c}} &
\colhead{Sky Sigma\tablenotemark{d}} &
\colhead{Upper Limit\tablenotemark{b}} &
\colhead{Sky\tablenotemark{c}} &
\colhead{Sky Sigma\tablenotemark{d}} \\
\colhead{} &
\colhead{} & 
\colhead{} &
\multicolumn{3}{c}{(10$^{32}$ \es)} &
\multicolumn{3}{c}{(10$^{32}$ \es)} &
\multicolumn{3}{c}{(10$^{32}$ \es)}
}

\startdata
 \cutinhead{\htrs}
None & 1386 & 1720 & 1.889 & 2.18 & 0.018 & 2.389 & 1.122 & 0.012 & 1.421 & 0.892 & 0.011
 \\ R $<$17 pc & 711 & 865 & 1.116 & 2.257 & 0.023 & 0.117 & 1.123 & 0.015 & 13.02 & 0.854 & 0.013
 \\ L$_{\rm{H}\alpha}>4\times10^{36}$ \es & 476 & 617 & 3.186 & 2.032 & 0.041 & 6.678 & 0.983 & 0.025 & 3.435 & 0.674 & 0.021
 \\ R $>$34 pc & 83 & 106 & 113.128 & 1.794 & 0.08 & 35.079 & 0.175 & 0.011 & 379.804 & 0.064 & 0.005
 \\ L$_{\rm{H}\alpha}>10^{35}$ \es & 1231 & 1515 & 1.3 & 2.16 & 0.02 & 2.087 & 1.104 & 0.013 & 1.121 & 0.872 & 0.012
 \\ L$_{\rm{H}\alpha}>4\times10^{36}$ \es\ \& R $<$17 pc & 132 & 172 & 105.83 & 1.442 & 0.061 & 63.978 & 0.131 & 0.007 & 28.521 & 0.04 & 0.003
 \\ \cutinhead{Star Clusters}
  None & 54 & 86 & 24.058 & 4.641 & 0.054 & 0.632 & 1.573 & 0.03 & 0.187 & 3.054 & 0.045
 \\ R$_{\rm{eff}}>$0.4 pc & 33 & 51 & 2.302 & 4.42 & 0.069 & 0.74 & 1.473 & 0.038 & 0.553 & 2.871 & 0.057
 \\ F475W $>$20.5 & 21 & 28 & 1.678 & 3.073 & 0.09 & 1.157 & 1.283 & 0.05 & 1.158 & 1.651 & 0.073
 \\ R$_{\rm{eff}}<$0.4 pc & 21 & 35 & 0.711 & 4.952 & 0.089 & 2.156 & 1.693 & 0.05 & 0.07 & 3.276 & 0.074
 \\ F475W $<$19 & 15 & 25 & 1.044 & 6.528 & 0.113 & 3.587 & 1.685 & 0.055 & 0.818 & 4.723 & 0.097
 \\ R$_{\rm{eff}}<$0.4 pc \& F475W $<$19 & 4 & 7 & 3.315 & 7.475 & 0.222 & 2.827 & 1.565 & 0.091 & 2.413 & 5.602 & 0.197
 \\ 
\enddata
\tablecomments{A foreground absorption term (assuming Galactic $N_{\rm H}\approx6.6\times10^{20}$ cm$^{-2}$ \citep{dickey-90} and a power-law spectrum of $\Gamma\sim1.7$) would increase all our luminosities by $\sim14\%$.}
\tablenotetext{a}{From ACIS-I and ACIS-S data.}
\tablenotetext{b}{On source aperture (radius of 1$\arcsec$ and 3$\arcsec$ for star clusters and \htrs\ respectively) luminosity.}
\tablenotetext{c}{The background aperture (outside a radius of 5$\arcsec$ and 10$\arcsec$ for star clusters and \htrs\ respectively) luminosity per pixel.}
\tablenotetext{d}{Uncertainty in the background per pixel, determined by averaging the upper and lower limits calculated using equations (9) and (14) of \citet{gehrels04-86}.}
\end{deluxetable*}
\end{turnpage}

\end{document}